# Denoising Interferometric Observations Using Visibility-Informed Neural Networks

Jason P. Terry [1,2,3]⋆, Cassandra Hall [2,3] Sergei Gleyzer [4],
[1] *Department of Earth Sciences, University of Oxford, Oxford, UK*
[2] *Department of Physics and Astronomy, The University of Georgia, Athens, GA 30602, USA*
[3] *Center for Simulational Physics, The University of Georgia, Athens, GA 30602, USA*
[4] *Department of Physics and Astronomy, The University of Alabama, Tuscaloosa, AL 35487, USA*

Accepted XXX. Received YYY; in original form ZZZ

**ABSTRACT**

The upcoming observations from the Square Kilometer Array Observatory will provide the astronomical community with a wealth of observations of important objects at long wavelengths. Full analysis of these outputs will necessitate specialized methods and software. Using synthetic observations of protoplanetary discs as an example, we present VIREO, a machine-learning-based visibility-informed method for denoising interferometric images. VIREO operates on image-plane observations, but it is informed by the interferometric measurement process through the UV-derived point spread function supplied as an additional input and used in the loss function. VIREO outperforms traditional cleaning methods and PSF-ignorant denoising models by producing data that is quantitatively cleaner and more conducive to analysis of the planets within the disc. Applying VIREO to archival ALMA data creates images with significantly less background noise, while maintaining, and in some cases enhancing, the substructure. By demonstrating the general utility of visibility-informed models, our results suggest that VIREO can be applied across interferometric observatories when trained on appropriate datasets.

## 1 INTRODUCTION

Interferometry is the state-of-the-art method for astronomical observations at long wavelengths. By combining the data from an array of telescopes, interferometry creates images with a resolution that far surpasses the abilities of any single telescope. Facilities, such as the Atacama Large Millimeter/Submillimeter Array (ALMA), have used interferometric methods to uncover an unprecedented wealth of new information of astronomical objects, such as protoplanetary discs, distant galaxies, and even directly imaging supermassive black holes ALMA Partnership et al. (2015a); Event Horizon Telescope Collaboration et al. (2019); Hodge & da Cunha (2020); Díaz Trigo et al. (2023). In the next few years, the next generation of array telescopes, such as the Square Kilometer Array Observatory (SKAO) and next-generation Very Large Array (ngVLA), will begin observations and further push the boundaries of astronomy.

In the meantime, it is essential to prepare for the data that these telescopes will produce. Combining the array observations and cleaning the results towards high-quality data analysis products is a difficult task that requires specialized methods and software. For example, `CASA` (CASA Team et al. 2022) is widely used software that, among other successes, has been instrumental in creating data products from large surveys, such as MAPS, DSHARP, and exoALMA (Öberg et al. 2021; Andrews et al. 2018; Teague et al. 2025), that have led to significant insights in the field of observing protoplanetary discs to understand planet formation. More recently, machine learning methods have been used to denoise interferometric data and used $u-v$ information to help localize sources (Taran et al. 2023; Drozdova et al. 2024). These works indicate that machine learning may prove to be a powerful tool in processing and analysing interferometric observations.

While these methods have led to undeniable successes and progress, previous work has shown that upcoming data from the SKAO may have significant noise levels that can inhibit the analysis of protoplanetary discs (Ilee et al. 2020). Observations over reasonable timescales, e.g., tens of hours, will lead to noisy images that can obscure substructure – one of the important features in discs. Capitalizing on SKAO's data is essential, but this requires clean observations that are conducive to detailed analysis of the most subtle features.

With this motivation, we put forth a new machine learning method for denoising SKAO observations that produces data products with low noise levels and detailed substructures. We introduce VIREO, a visibility-informed reconstruction method for enhanced observations in the image-domain. VIREO denoises continuum image-domain observations by conditioning the model on both the observed image and the associated PSF. Here, "visibility-informed" refers to the fact that the PSF encodes information about the interferometric UV sampling and imaging choices used to produce the observation; VIREO does not ingest complex visibilities or operate directly in the visibility domain. This approach gives the model additional information about the observational qualities that produce the noise and resolution in the observation. Compared to our experiments with both traditional methods and denoising models that ignore PSF, this method leads to higher-quality outputs. We demonstrate its utility on synthetic observations created from simulations of protoplanetary discs and archival ALMA data in order to test it on existing observations. This method will provide a valuable tool for analyzing SKAO data, while the general approach of visibility-informed denoising models can be extended to all interferometric observatories.

⋆ E-mail: jason.terry@earth.ox.ac.uk

In this paper, we first introduce the methods to create the data used for training and testing the VIREO model in Section 2. We present example denoising results, comparisons with other methods, image metrics, and ALMA applications in Section 3. In Section 4, we discuss the results and implications of our method, and we offer our conclusions in Section 5. Further denoising results are shown in Appendix B and Appendix C.

## 2 METHODS

### 2.1 Data Preparation

#### 2.1.1 *Simulations*

We use 1,489 smoothed-particle hydrodynamics (SPH) simulations of gas-only discs with $10^6$ SPH particles to train and test the denoising models. While including dust would have been ideal, computational limits make simulating so many dusty discs impractical. The simulations are made using the same parameters given in Terry et al. (2022). The discs are simulated using the publicly available code `PHANTOM` (Price et al. 2018), and outputs are passed through `MCFOST` (Pinte et al. 2006, 2009) radiative transfer calculations to generate the baseline image-plane synthetic observation. The system and observational parameters, including disc size, disc mass, stellar mass, stellar age, number of planets, mass of planets, locations of planets, disc viscosity, and sky orientation, are sampled across a Latin Hypercube (McKay et al. 1979) to ensure the highest possible degree of coverage of parameter space. All discs contain between 0 and 4 planets. For more details on the simulations and parameters of the discs, see Terry et al. (2022).

#### 2.1.2 *Synthetic Observations*

The SPH results are used as inputs into the radiative transfer code `MCFOST` (Pinte et al. 2006, 2009). The stellar parameters (temperature and radius) are taken from Siess et al. (2000) isochrones using uniformly sampled ages between 1 Myr and 5 Myr. Each disc's inclination, azimuth, and density are taken from the original Latin Hypercube used to create the simulations. $10^8$ photon packets are used. Because larger grains are more visible using SKAO-Mid, we assumed a logarithmic dust-size distribution between 0.1$\mu$m$\leq a \leq$ 1cm split into 100 bins, with a dust-to-gas ratio of 1:100. We imposed a power-law between grain size and number of grains, $dn(a) \propto a^{-3.5}da$. Grains were assumed to be carbon and silicates (Draine & Lee 1984). Planets were turned off as blackbody sources.

The outputs from `MCFOST` are clean image-plane radiative-transfer models at 12.5 GHz with resolutions of 1 au per pixel (600×600 pixels) and are considered the ground truth for our models, i.e., the standard against which denoising metrics are measured during training and evaluation.

We use the clean image-plane data as inputs to create noisy 12.5 GHz synthetic observations designed to approximate SKAO-Mid restored image products. Figure 1 shows the array configuration we used (left) and an example UV coverage map (right) over a 10 hour observation with 10 second exposures. The array consists of 133 dishes with a maximum baseline of 154 km and is set at a latitude of -30°42'46.5294" S. Ilee et al. (2020) previously analysed the noise and resolution of 12.5 GHz SKAO-Mid observations of protoplanetary discs. Their results are given in Table 1. We adopt their calculated parameters to create our observations.

Additional parameters are necessary to fully calculate the synthetic observations. We sampled the single exposure time, i.e., seconds per frame, uniformly between 10 and 30 seconds. The target declination was sampled uniformly across the visible sky at the SKAO-Mid location with a minimum elevation of 20°. The beam's axial ratio is randomly sampled uniformly between 0.5 and 1. We also sample the peak SNR and $\sigma_{\rm RMS}$ between 80% and 120% of the values given in Table 1 to create further variability between observations without qualitatively leaving the results from Ilee et al. (2020). We do not uniformly distribute the observation classes, i.e., integration time and resolution, as shown in Table 1. Instead, the majority of the dataset has intermediate-level noise. The smallest fraction (5%) of the data is reserved for cleanest synthetic observations (67 mas, 1000 hours) because these provide the model with the least amount of noise information and real observations would rarely, if ever, have such long integrations.

The UV sampling is generated from the SKAO-Mid antenna coordinates. For a source at the sampled declination, we compute the projected baselines over the allowed hour-angle range for the selected total integration time and exposure time. The projected baselines are converted to $u, v$ coordinates in units of wavelengths at 12.5 GHz and accumulated onto the Fourier grid corresponding to the image size and pixel scale. We add the conjugate UV samples, apply a small Gaussian smoothing kernel in the UV plane, and enforce Hermitian symmetry. The PSF is then computed as the normalized inverse Fourier transform of this gridded UV sampling function.

We emphasize that this dirty beam is not used as the fiducial convolution kernel for the synthetic image-plane inputs. Instead, our noisy restored image-plane observations are designed to approximate restored continuum science products, of the type commonly used for downstream image-domain analysis, rather than raw undeconvolved dirty maps. Real science products may additionally include primary-beam correction, whose spatially varying noise is not included in our fiducial approximation.

To construct the Gaussian restoring beam, we use the observational setup to set the target angular resolution, $\theta_{\rm targ}$, and sample an axial ratio $q = B_{\rm min}/B_{\rm maj}$ uniformly between 0.5 and 1. We define $\theta_{\rm targ}$ as the geometric mean of the restoring-beam axes, so that

$$B_{\rm maj} = \theta_{\rm targ}/\sqrt{q}, \qquad B_{\rm min} = \theta_{\rm targ}\sqrt{q}.$$

The beam position angle is measured from an elliptical Gaussian fit to the central core of the UV-derived dirty beam. Thus, the restoring beam is not a fit to the full dirty beam with enforced properties; rather, it is a prescribed Gaussian clean/restoring beam whose angular resolution and axial ratio are set by the observing setup and whose orientation is informed by the UV-derived PSF.

The clean radiative-transfer image is convolved with this Gaussian restoring beam to produce the noiseless restored-image model. This choice is intentional: VIREO is trained and applied to restored image-plane science products, not to raw dirty maps or directly to complex visibilities. A direct convolution with the full dirty beam would instead approximate an undeconvolved dirty map, including the sidelobe response of the incomplete UV sampling, which is not the target data product of this work.

Finally, correlated residual noise is generated by convolving image-plane white noise with the same Gaussian restoring beam.

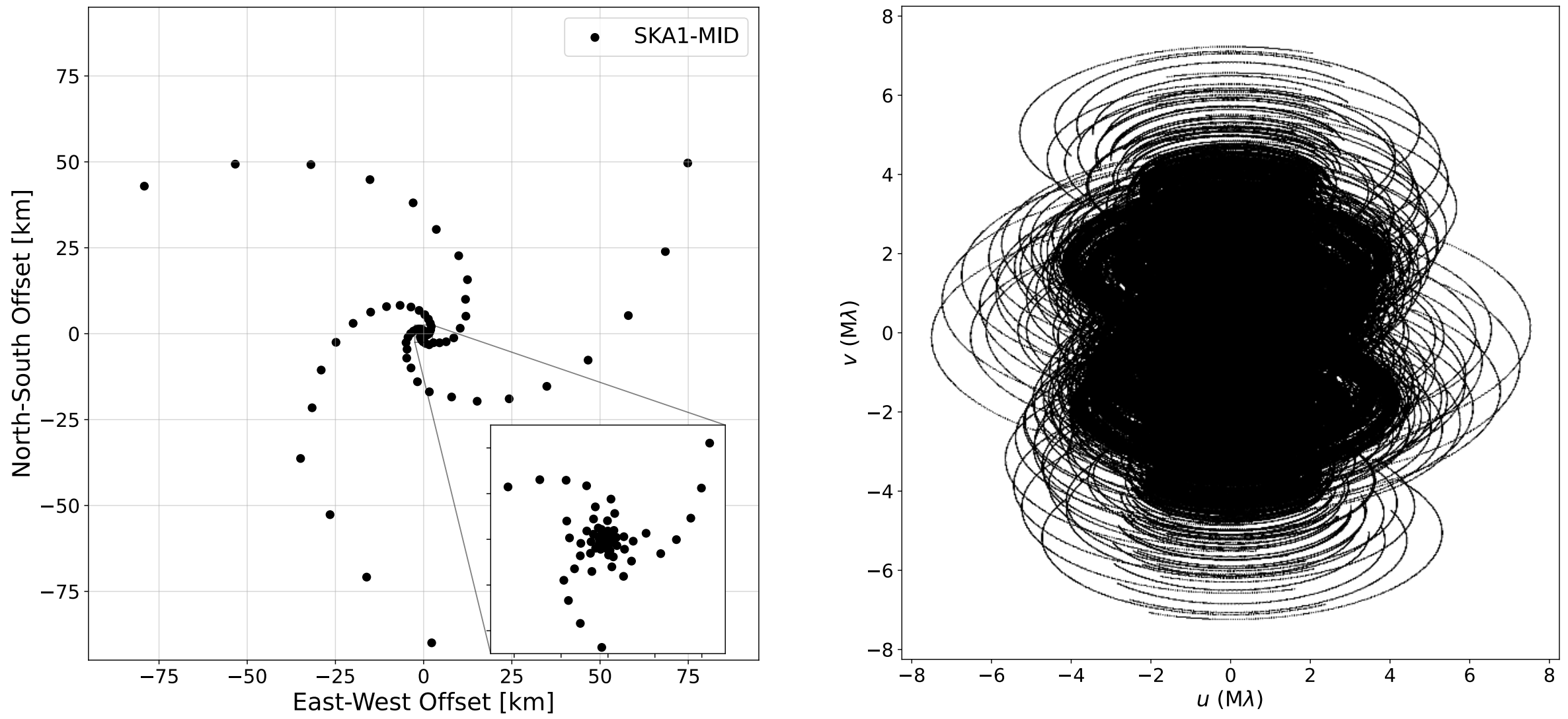

**Figure 1.** SKAO-Mid array configuration and UV coverage map. Left: SKAO-Mid configuration. The array consists of 133 dishes with a maximum baseline of 154 km. The inset shows the inner 5x5 km of the array. Right: UV coverage from a 10 hour observation with 30 second exposures.

This imposes beam-scale correlations on the residual field and approximates the noise properties of restored continuum images associated with the RMS values adopted from Ilee et al. (2020). The raw noise realization is then scaled so that its off-source RMS matches the sampled target RMS. The beam-convolved signal model is scaled so that its peak intensity equals the sampled peak SNR multiplied by the measured noise RMS. The final noisy restored synthetic observation is therefore

$$I_{\mathrm{obs}} = I_{\mathrm{model}} + n,$$

where $I_{\mathrm{model}}$ is the scaled restoring-beam-convolved signal model and $n$ is the scaled beam-correlated residual-noise realization. Figure 2 and Table 1 give the overall data creation pipeline and parameters, respectively. We show example noisy restored outputs and the PSFs from an example image-plane synthetic observation (Figure 3) for the 34 mas and 67 mas resolution in Figure 4 and Figure 5, respectively.

This image-plane procedure is much faster than generating full measurement sets and running an iterative imaging pipeline for every synthetic observation. Instead, it uses a UV-derived PSF together with restored-image resolution and noise properties and the RMS/SNR values informed by Ilee et al. (2020). We therefore use this approximation to generate the large training set.

The fiducial synthetic observations do not reproduce every artifact that may remain in a real restored interferometric image. In particular, they do not explicitly include residual dirty-beam sidelobes, imperfect deconvolution or calibration residuals, or the spatially varying noise introduced by primary-beam correction. These limitations affect the absolute realism of the synthetic data. However, the principal controlled experiment in this work compares PSF-informed and PSF-ignorant models trained and evaluated on the same images and clean targets. The comparison therefore remains informative about whether conditioning an image-domain model on the PSF improves reconstruction within the restored-image data-product regime considered here.

We compare our denoising methods with traditional cleaning using `CASA` (CASA Team et al. 2022). We use the synthetic observed image and corresponding PSF as inputs to a multiscale deconvolution pipeline. The PSF core is fit with an elliptical Gaussian to define the restoring beam, an SNR-based mask is constructed, and CASA multiscale deconvolution is run with the scales and thresholds listed in Table A1. This provides a representative CASA multiscale reconstruction baseline using the same synthetic image and associated PSF. Because the input is an idealized restored-image approximation rather than a full measurement-set simulation, this comparison is not intended as an end-to-end benchmark of standard interferometric imaging. While we recognize that `CASA` has a variety of methods and parameters that can be tuned for improved performance on specific observations, we consider our method to be representative, although not exhaustive.

Figure 6 shows the overall pipeline for our model development and comparisons.

### 2.2 Machine Learning-based Denoising and Analysis Models

We implement 3 different machine learning models in this work. The first model denoises the noisy synthetic observations using both the observation and PSF as inputs. The second model also denoises, but the PSF is never used. The third machine learning model learns the properties of the disc, i.e., are there any planets (classification) and how many planets (regression). All the models are built using `PyTorch Lightning` (Falcon & The PyTorch Lightning team 2019), a wrapper to facilitate `PyTorch` (Paszke et al. 2019) implementations.

#### *2.2.1 Denoising Models*

The synthetic observations provide two image-like inputs: the noisy synthetic observation and the PSF. Both the clean and noisy observations are normalized such that all pixels fall between -1 and 1, and the PSF is normalized such that its peak is 1 and located at the image's center. Both of these are the same shape: 600×600 pixels.

We use these in a U-Net (Ronneberger et al. 2015) model to output a recreation of the clean data synthetic observation that was used to create the noisy data. U-Nets are powerful encoding-decoding convolutional networks that include skip connections between upper layers to preserve high-level features of the inputs while also allowing the network to compress the information into a lower-dimensional latent space. The U-Net has three down-sampling layers, using a combination of residual (He et al. 2016) and inception (Szegedy et al. 2014) blocks that use DropBlock (Ghiasi et al. 2018). A 76x76x128 bottleneck with global multihead self-attention (Cordonnier et al. 2020) (2 heads) leads into three up-sampling layers that mirror the down-sampling layers. The final up-sampling layer has sliding window self-attention (Liu et al. 2021a). As typically for U-Nets, the output of each down-sampling layer is also used as an additional input for the corresponding up-sampling layer. After the final up-sampling layer, a small Laplacian filter correction is added for edge enhancement. The resulting image is passed through a depth-wise convolution followed by a 1x1 convolution to help smooth rough edges. Finally, the edge-smoothed output is passed through a learnable *asinh* activation ($k\,asinh(x/k)$, where $k$ is a learnable parameter) to approximately put the output in the original data's range without causing the saturation that could arise from a *tanh* activa-tion. The activated output is treated as the corrective term, i.e., the difference between the clean target and noisy input, and is added to the original noisy input to recreate the clean target. The network output is converted back to physical image units by inverting the normalization applied to the input image. Thus, the flux metrics are computed after converting the reconstruction back to the original image units, not in normalized units. Explicitly, for each sample, the observed image is mapped to the interval $[-1, 1]$ using the stored noisy-image normalization parameters $a = I_{\rm obs,min}$ and $b = I_{\rm obs,max}$. Therefore, $I_{{\rm obs},norm} = 2\frac{I_{\rm obs}-a}{b-a} - 1$. The network predicts a correction in normalized units, $\Delta_{\rm norm}$, which is added to the normalized observed image: $\hat{I}_{\rm norm} = I_{\rm obs,norm} + \Delta_{\rm norm}$. Before computing fluxes and image-domain metrics in physical units, we apply the inverse transform, $\hat{I} = a + \frac{\hat{I}_{\rm norm}+1}{2}(b-a)$.

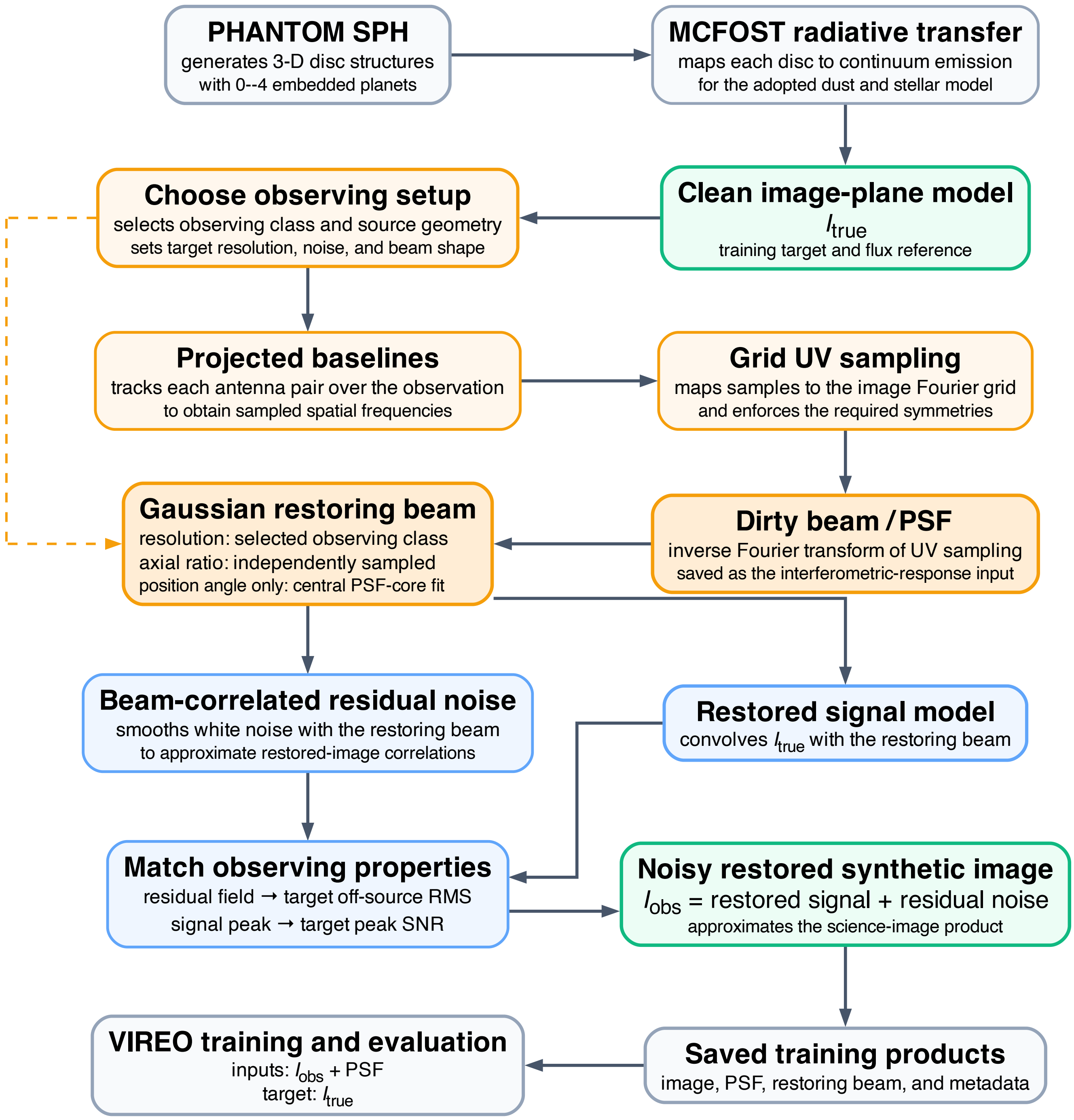

**Figure 2.** Workflow for generating the synthetic SKAO-Mid observations used to train and evaluate VIREO. Hydrodynamical simulations are post-processed with MCFOST to produce clean image-plane radiative-transfer models. For each clean image, we sample an observing setup and compute the UV sampling from the SKAO-Mid antenna coordinates, source declination, and observing track. The gridded UV sampling is Fourier transformed to obtain the UV-derived dirty beam/PSF, which is saved and supplied to VIREO. The image-plane input is a noisy restored-image approximation. We construct a Gaussian restoring beam whose target angular resolution is set by the observing class, whose axial ratio is sampled, and whose position angle is measured from the central core of the UV-derived dirty beam. The clean image is convolved with this restoring beam, beam-correlated residual noise is generated and scaled to the target RMS, and the signal is scaled to the target peak SNR. VIREO takes the noisy restored image and PSF as inputs and uses the clean image-plane model as the training target.

**Table 1.** Parameters used in the synthetic-observation workflow shown in Figure 2. Stage names follow the terminology in that figure. The lower panel lists the six SKAO-Mid observing classes adopted from Ilee et al. (2020); $\theta_{\rm targ}$ is the target angular resolution, $t$ is the total integration time, $\sigma_c$ is the continuum sensitivity, $\sigma_\nu$ is the visibility noise, and $f_{\rm data}$ is the fraction of synthetic observations assigned to each class. For the ALMA-like synthetic observations, the hydrodynamical models are unchanged, while the radiative-transfer wavelength and grain-size prescription and the observing-specific array, resolution, RMS, peak-SNR, and declination distributions are replaced by their ALMA counterparts.

| **Stage** | **Parameter** | **Value or sampling prescription** |
|---|---|---|
| PHANTOM SPH | Hydrodynamical models | $10^6$ gas particles; 0–4 planets; stellar mass, disc mass, disc size, viscosity, and planet masses/radii sampled with the Latin hypercube of Terry et al. (2022). |
| MCFOST radiative transfer | Image calculation<br>Dust and stellar prescription | MCFOST; $10^8$ photon packets; 12.5 GHz; 600 × 600 pixels; 1 au pixel$^{-1}$.<br>Dust-to-gas ratio 1:100; 0.1 $\mu$m $\leq a \leq$ 1 cm in 100 logarithmic bins; $dn(a) \propto a^{-3.5}da$; stellar age uniform from 1–5 Myr with temperature/radius from Siess et al. (2000). |
| Choose observing setup | Observing class<br>Per-sample geometry and cadence<br>Beam/noise variation | One of the six classes in the lower panel, which sets $\theta_{\rm targ}$, $t$, $\sigma_{\rm RMS}$, peak SNR, and $f_{\rm data}$.<br>Integration time uniform from 10–30 s; declination uniform in visible sky area subject to a minimum elevation of 20°.<br>Axial ratio $q = B_{\rm min}/B_{\rm maj}$ uniform from 0.5–1; RMS and peak SNR sampled within ±20% of the class values. |
| Projected baselines | Array and track | 133 SKAO-Mid dishes; maximum baseline 154 km; site latitude −30°42′46.5294″; antenna-pair baselines projected over the allowed hour-angle range at 12.5 GHz. |
| Grid UV sampling | Fourier grid | Projected samples accumulated on the image Fourier grid; conjugates added; class-dependent Gaussian UV smoothing applied; Hermitian symmetry enforced. |
| Dirty beam / PSF | Construction | Normalized inverse Fourier transform of the gridded UV sampling; peak centered and normalized to unity. |
| Gaussian restoring beam | Size and shape<br>Position angle | $B_{\rm maj} = \theta_{\rm targ}/\sqrt{q}$ and $B_{\rm min} = \theta_{\rm targ}\sqrt{q}$, so $\theta_{\rm targ} = \sqrt{B_{\rm maj}B_{\rm min}}$.<br>Measured from an elliptical-Gaussian fit to the central core of the UV-derived PSF; only the position angle is taken from this fit. |
| Restored signal model | Signal convolution | Clean radiative-transfer image convolved with the Gaussian restoring beam. |
| Beam-correlated residual noise | Residual model | Image-plane white noise convolved with the same restoring beam to impose beam-scale correlations. |
| Match observing properties | Scaling | Residual field scaled to the sampled off-source RMS; beam-convolved signal scaled so that its peak equals the sampled peak SNR multiplied by the measured noise RMS. |
| Saved training products | FITS contents | Noisy restored image, UV-derived PSF, Gaussian restoring beam, and beam/RMS/SNR metadata. |
| VIREO training and evaluation | Inputs and target | Inputs: noisy restored image and PSF; target: clean image-plane radiative-transfer model. |

| $\theta_{\rm targ}$ (mas) | $t$ (h) | $\sigma_c$ ($\mu$Jy beam$^{-1}$) | $\sigma_\nu$ (mJy) | $\sigma_{\rm RMS}$ ($\mu$Jy beam$^{-1}$) | Peak SNR | $f_{\rm data}$ |
|---|---|---|---|---|---|---|
| 34 | 10 | 2.4 | 0.56 | 0.83 | 4.90 | 0.10 |
| 34 | 100 | 2.4 | 0.56 | 0.26 | 7.73 | 0.25 |
| 34 | 1000 | 2.4 | 0.56 | 0.08 | 16.80 | 0.25 |
| 67 | 10 | 1.2 | 0.28 | 0.44 | 7.58 | 0.10 |
| 67 | 100 | 1.2 | 0.28 | 0.14 | 17.17 | 0.25 |
| 67 | 1000 | 1.2 | 0.28 | 0.05 | 44.75 | 0.05 |

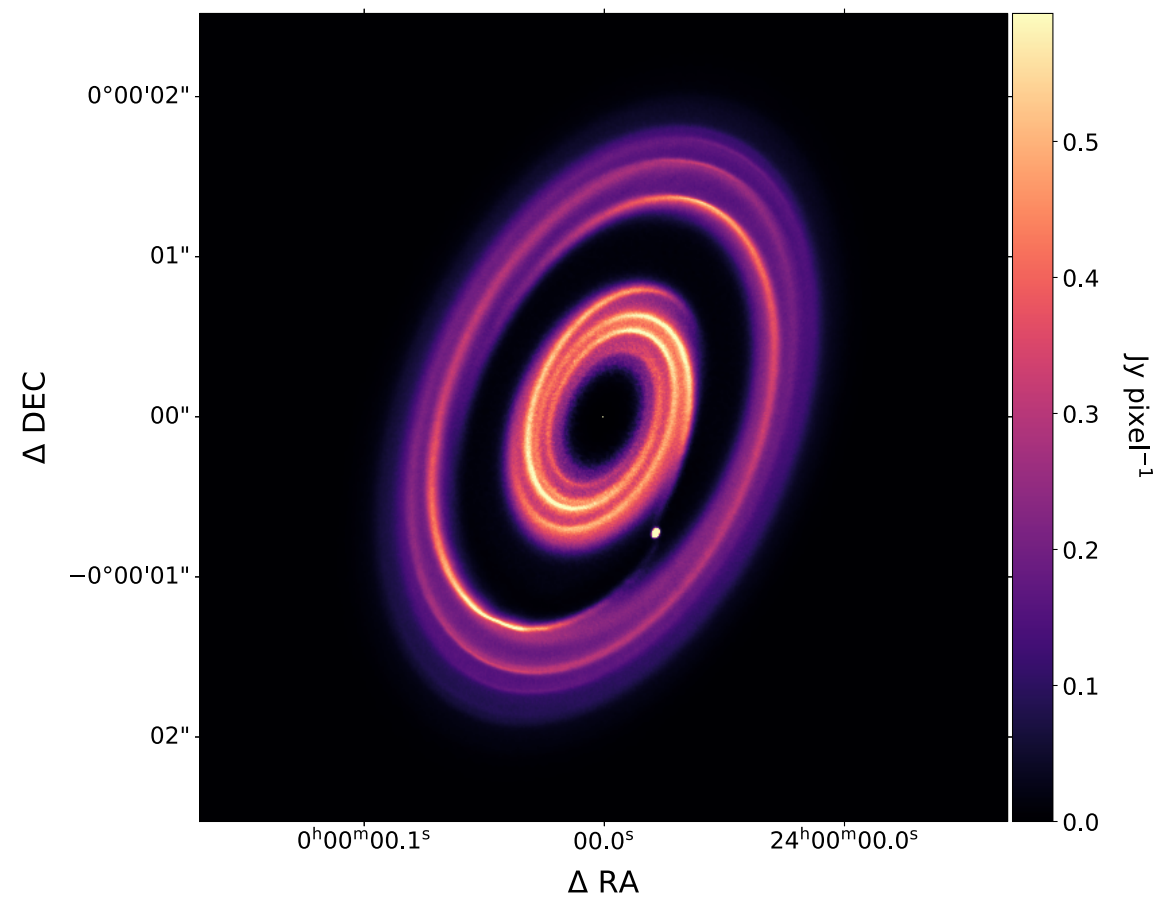

**Figure 3.** Example image-plane synthetic observation for a disc containing 3 planets.

The overall architecture is given in Figure 7, and a more detailed view can be seen in Figure D1 in Appendix D.

The key addition to this model is the introduction of feature-wise linear modulation layers (FiLM) (Perez et al. 2017). These layers are added to all up-sampling layers and the bottleneck. The purpose of the layers is to inform the network on the observation's PSF, thereby giving it crucial information about the beam and other observational parameters that resulted in the noisy image's blurring and noise. The FiLM layers are only added to the decoding layers and bottleneck because these layers are most responsible for reconstructing the data whereas the encoder is designed to compress the image input into a smaller latent space and extract important features. Preliminary experiments showed that adding FiLM layers to the encoder had negligible improvements on model performance while adding a significant number of parameters. We therefore only apply the FiLM layers to the decoding and bottleneck parts of the model to directly inform the reconstructions.

All FiLM layers rely on a context vector that is extracted from the PSF itself. The context is created in two parts. First, the geometric features of the beam (major FWHM, minor FWHM, eccentricity, cosine and sine of position angle, and the pixels-per-beam) are directly extracted from the PSF. This gives the model direct access to the beam's information without need of user intervention. Separately, the PSF itself is passed through a convolutional encoder. This output is concatenated with the extracted features and passed through another series of convolutions. This creates the shared context vector, $z$, that is used for the bottleneck's FiLM layer. The other context vectors are made by passing the shared vector through separate output heads.

For each FiLM layer, the corresponding context vector is used as an input into a standard linear layer to produce two modulating vectors, $\gamma$ and $\beta$. These modulating vectors are added to the output, $y$, from the original convolution block, e.g., a residual + inception block (optionally some attention treatment), that uses the output, $x$, from the previous FiLM layer to produce the modulated output, $y_{\rm mod} = y_n(1+\gamma)+\beta$ (where $y_n = {\rm GroupNorm}(y)$). A learnable gate parameter, $g$, is then used to combine the newly modulated output with the original output. The FilM layer's final output, $y_{\rm out} = y + g(y_{\rm mod} - y_n)$, is then passed to the next block. This architecture allows the bottleneck and each up-sampling block to selectively include information from the observation's PSF. The learnable gate parameter gives the layer the ability to increase or decrease the importance of the PSF at a given depth and pass that information on through the rest of the network. Figure 8 gives an overview of the FiLM layers. A more detailed version can be found in Appendix D (Figure D2)

We train the model for 300 epochs with an Adam optimizer (Kingma 2014), allowing early stopping based on validation loss with a patience of 25 epochs. We reserve 20% of the data for testing and use 20% of the remaining data for validation. All weights are initialized according to (He et al. 2015). The main loss function is a combination of the reconstruction mean absolute error (MAE) and the multi-scale structural similarity index measure (MS-SSIM), where MS-SSIM is weighted by 0.5 compared to MAE. We also use an $l1$-starlet loss (Starck et al. 2007; Ajani et al. 2021) with a weight of 0.25. This brings out fine substructures like spiral wakes, disc-wide features like rings, and the overall global morphology of the disc instead of just focusing on the average brightness (MAE) and statistical similarity (MS-SSIM). This is important because the model could otherwise simply remove off-disc background noise to closely match the synthetic observation's average brightness and statistical measures with little regard to the importance and nuances of fine substructures. Finally, we explicitly include the PSF in the loss by computing the MAE between the clean image convolved with the PSF and the recreated image convolved with the PSF (weighted by 0.1). This leads to an overall loss of

$$\mathcal{L} = L_{\rm MAE} + 0.5\,[1 - \mathcal{L}_{\rm MS-SSIM}] + 0.25\mathcal{L}_{\rm Starlet} + 0.1\mathcal{L}_{\rm PSF} \quad (1)$$

We do a soft linear warm up of the learning rate over the first 5% of total steps, starting from 0.1% of the final learning rate ($10^{-4}$). After this warm up, we do a cosine annealing of the learning rate and also allow the learning to further drop by a factor of 1/2 if the validation loss does not improve over 5 epochs.

We also train an identical model without FiLM layers and PSF convolution loss. This PSF-ignorant model gives us the ability to quantify the value of including the PSF when denoising observations.

### 2.2.2 *Analysis Model*

In addition to the loss metrics of the denoising results themselves, we also measure the denoising's performance by comparing the performance of a downstream classification and regression models trained on the clean data. This is included as a proxy for the scientific value of denoising for analysis applications; denoising is of little use if there is no improvement in the downstream utility of the resulting images. We use a Swin V2 Vision Transformer (Liu et al. 2021b) backbone to jointly train a classification head (no planets or at least one planet) and a regression head (how many planets in the disc). The backbone is implemented using the pre-trained Swin V2-t model from `Torchvision` (maintainers & contributors 2016). Due to the class imbalance between (1/3 of the dataset has no planets whereas 2/3 of the dataset has at least 1 planet), we use focal loss (Lin et al. 2017) for the classifier. Focal loss is a modified cross-entropy loss that down-weights well-classified examples and places greater weight on difficult or minority-class examples, making it useful for imbalanced classification problems (Lin et al. 2017). The regression head outputs a density map whose sum is taken to the predicted number of planets. For the regression confusion matrices, this continuous prediction is rounded to the nearest integer and clipped to the allowed range. We use MAE with a TV regularizer (Estrela et al. 2016) as the loss function for the regression head. The model is first trained for 10 epochs using only the classification, after which the regression head is introduced, and the model is trained until the validation area under the receiver operating characteristic curve (AUC) does not improve for 25 epochs.

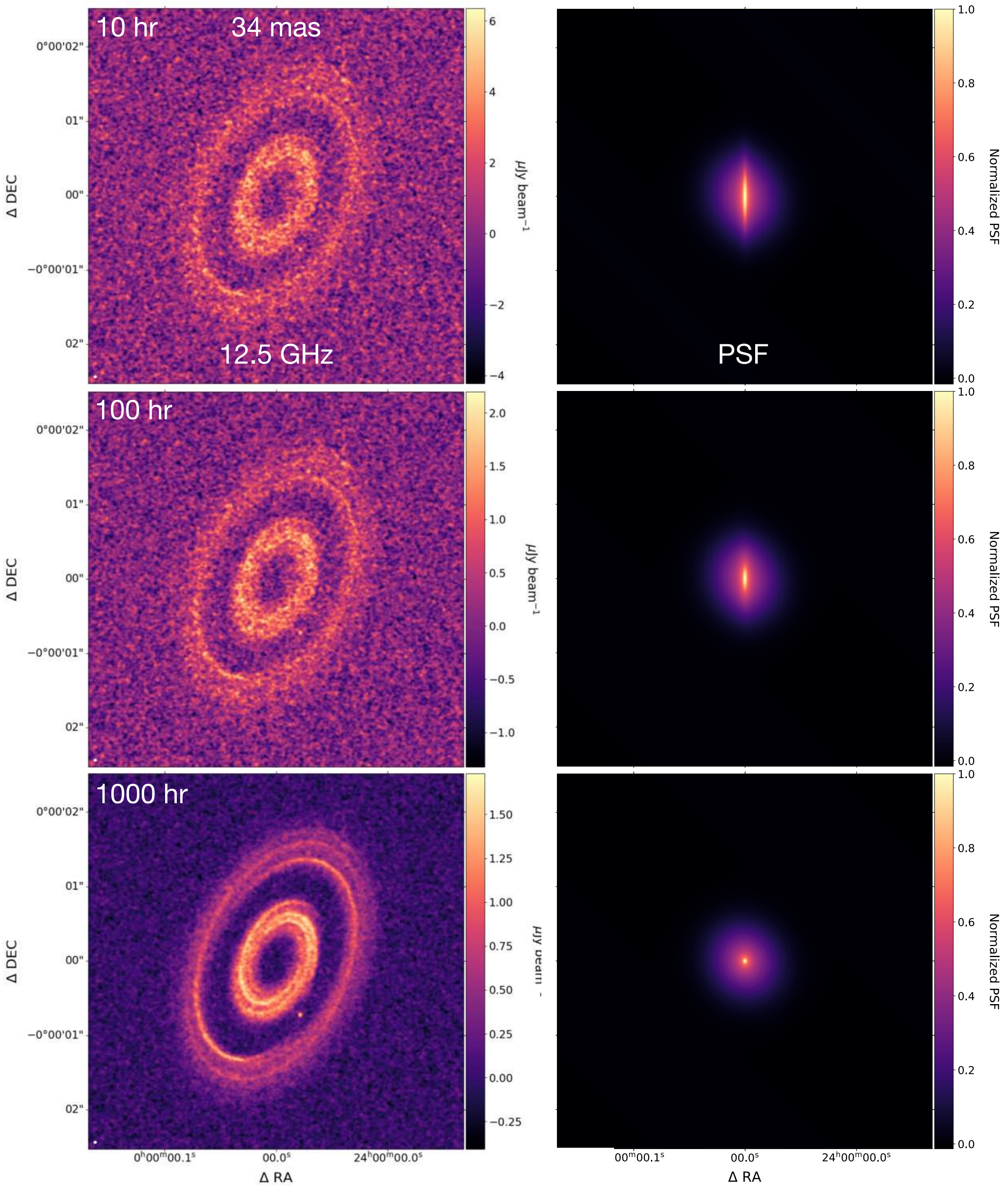

**Figure 4.** Synthetic noisy observations and point spread functions of the Figure 3 disc with 34 mas resolution. Left: 12.5 GHz observations. Right: corresponding PSFs. Top: 10 hour integration. Middle: 100 hour integration. Bottom: 1000 hour integration. As the integration time increases (top to bottom), the substructures become more clear, the noise decreases, and the PSF coverage becomes more complete.

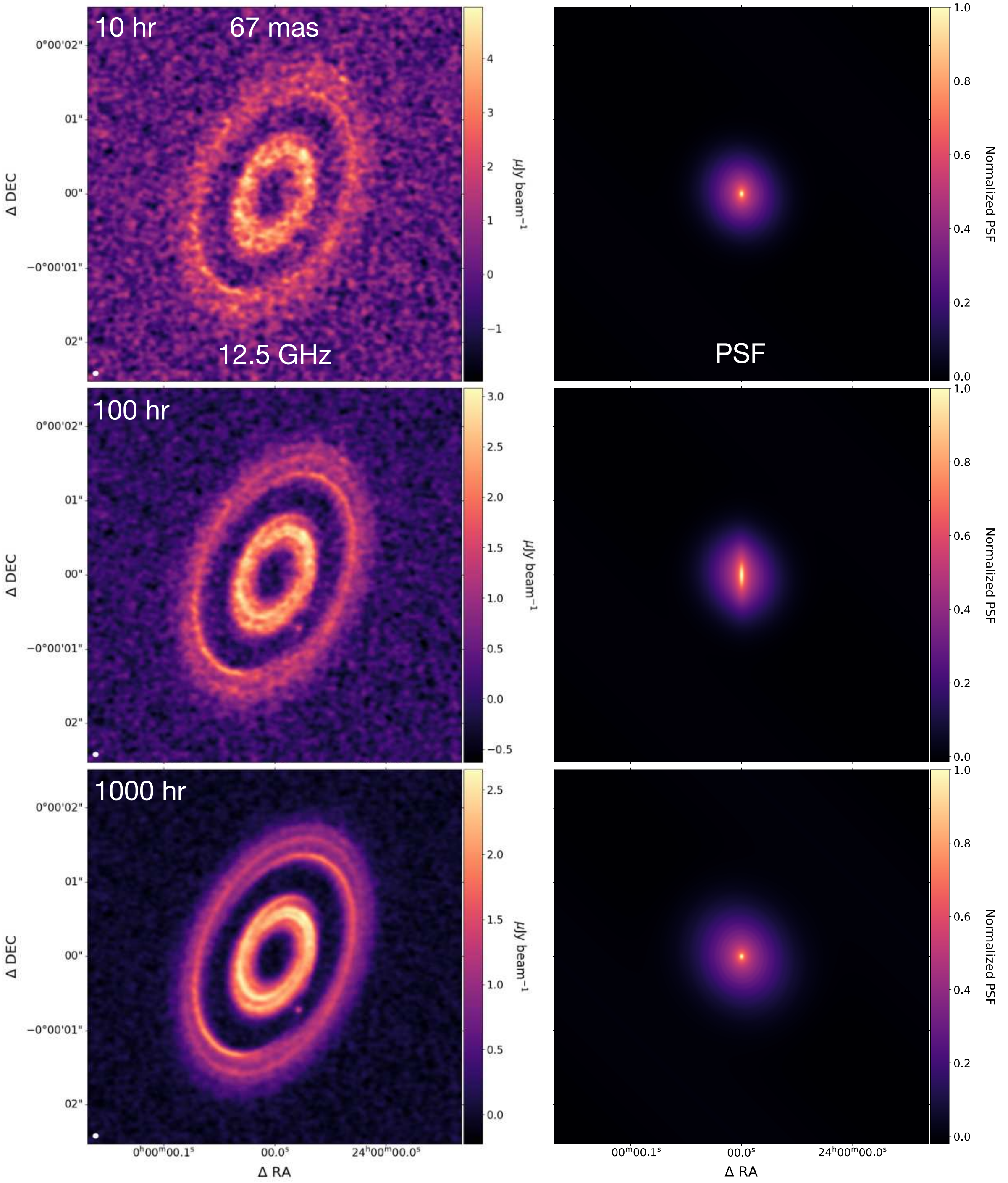

**Figure 5.** Same as Figure 4 except for 67 mas resolution. In all cases, the substructure is clearer and the observation less noisy than the corresponding observations in Figure 4.

The planet and no-planet simulations and synthetic observations are identical in all senses other than the fact that planet-hosting discs have $N_{\rm planet} > 0$. While a disc without a planet generally does not exhibit visible substructure, effects such as the addition of noise and projection may create fluctuations on the scale of planet-induced substructure, particularly for discs with few planets or low-mass planets.

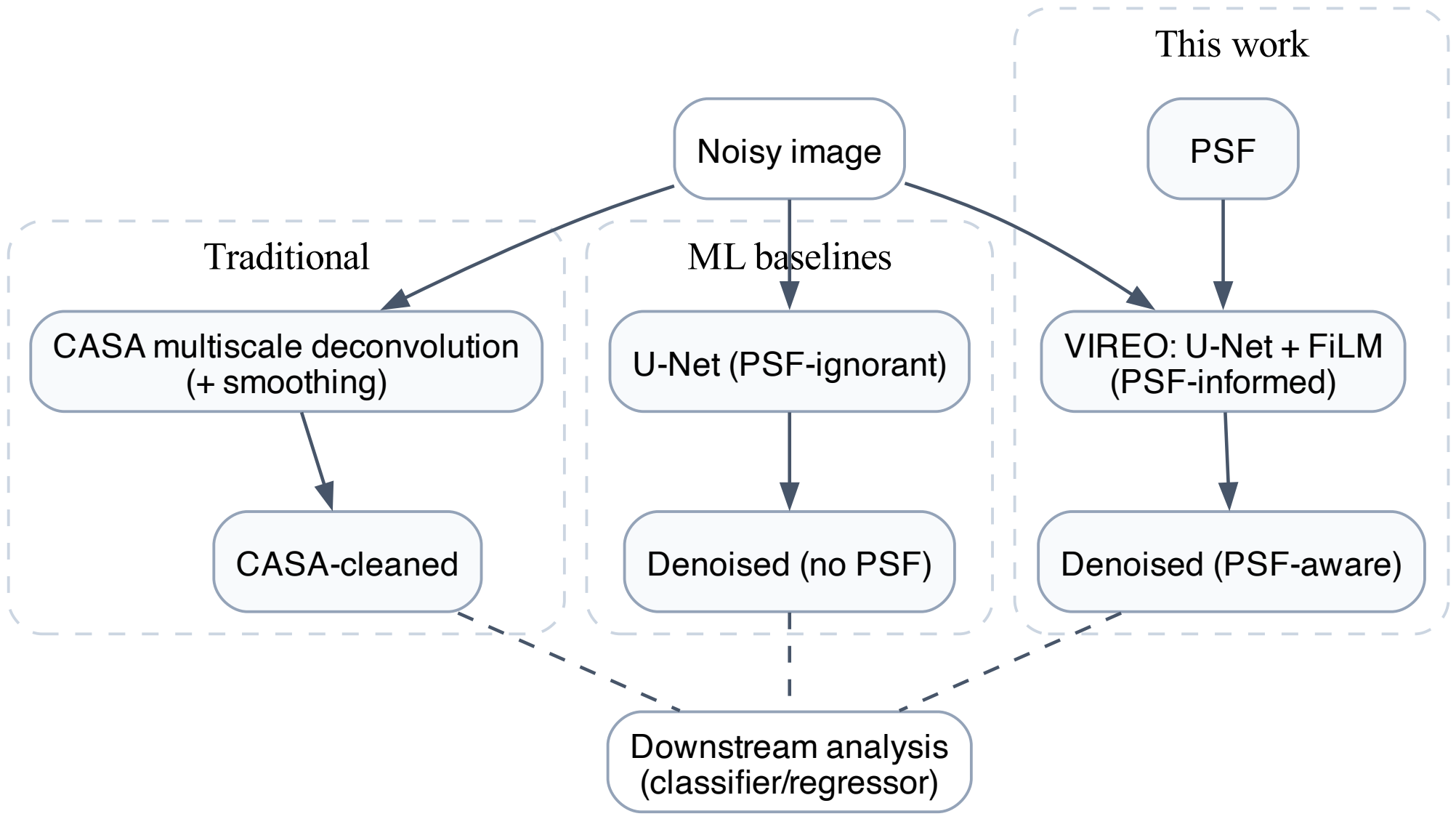

**Figure 6.** Deployment strategy for comparing VIREO and other techniques.

## 3 RESULTS

### 3.1 Denoising results

Results from VIREO show a clear, significant improvement in image clarity, substructure prominence, and noise levels. Figure 9 shows the outputs using the data in Figure 4 and Figure 5. In all cases, background noise is almost entirely removed. Substructure that was nearly invisible in the shorter integration and higher-noise observations becomes clear. For example, the outer spiral wake, as well as the rings themselves, in the 10 hour cases (top right of Figure 4 and Figure 5) are barely visible in the noisy image but are recovered in the VIREO reconstruction. As expected, there is an increase in output image quality that coincides with the quality of the input synthetic observation, but all denoised images show marked improvement. For further examples of VIREO outputs across all integration times and resolutions, see Appendix B.

We also compare the results of VIREO with those from traditional CASA cleaning methods, as described in Section 2.1.2. Because our fiducial synthetic inputs are restored-image approximations rather than full measurement-set simulations, this CASA comparison should not be interpreted as an exhaustive end-to-end interferometric imaging benchmark. Instead, it provides a traditional image-reconstruction baseline using the same synthetic image and associated UV-derived PSF. These results are shown in Figure 10. While CASA reduces noise and highlights substructure when compared with the noisy input image, the corresponding VIREO output more closely resembles the clean data and has even less noise and clearer substructure. Figure 11 shows the corresponding deprojected azimuthally averaged radial intensity profiles, and, once again, VIREO has the lowest error and preserves the peaks and gaps seen in the clean data. As seen in Figure 10 and Table 3, CASA has a wider range of predicted fluxes than either denoising models. It leads to both the highest and lowest fluxes, depending on if the pixel is on or off the disc, respectively. We consider this to be an artifact of the multiscale deconvolution passes. We deconvolve on scales to 8 times the beam size, so the signal is smeared across a larger area of the disc, which drives the resulting flux to higher levels. During our development, we found that multiscale cleaning led to clearer substructures and less noise, but it seems to have come at the cost of less well-preserved flux.

**Table 2.** Flux recovery metrics for different datasets.

| Dataset | Median flux error | Flux error MAD |
|---|---|---|
| Noisy data | -0.486 | 0.142 |
| VIREO cleaned | -0.017 | 0.019 |
| PSF-ignorant cleaned | 0.009 | 0.035 |
| CASA restored | -0.518 | 0.212 |

We quantify the results of VIREO, an identical model without any information from the PSF, and CASA-cleaned data in Figure 12. This figure shows that, for the metrics considered, VIREO more closely resembles the clean target data, compared to the other methods. CASA results in improved images, but both U-Nets lead to better results according to most metrics. While the PSF-ignorant model still creates high-quality denoised images, the inclusion of the PSF through FiLM layers and convolution loss leads to the best results.

We also demonstrate that VIREO improves important observational metrics. Figure 13 shows the off-source corner RMS for each dataset and subset of observational parameters. This is calculated using the standard deviation of the intensities of 10×10 boxes in each of the signal-free four corners of the observations (without reference to the ground-truth). CASA and the PSF-ignorant model lead to similar

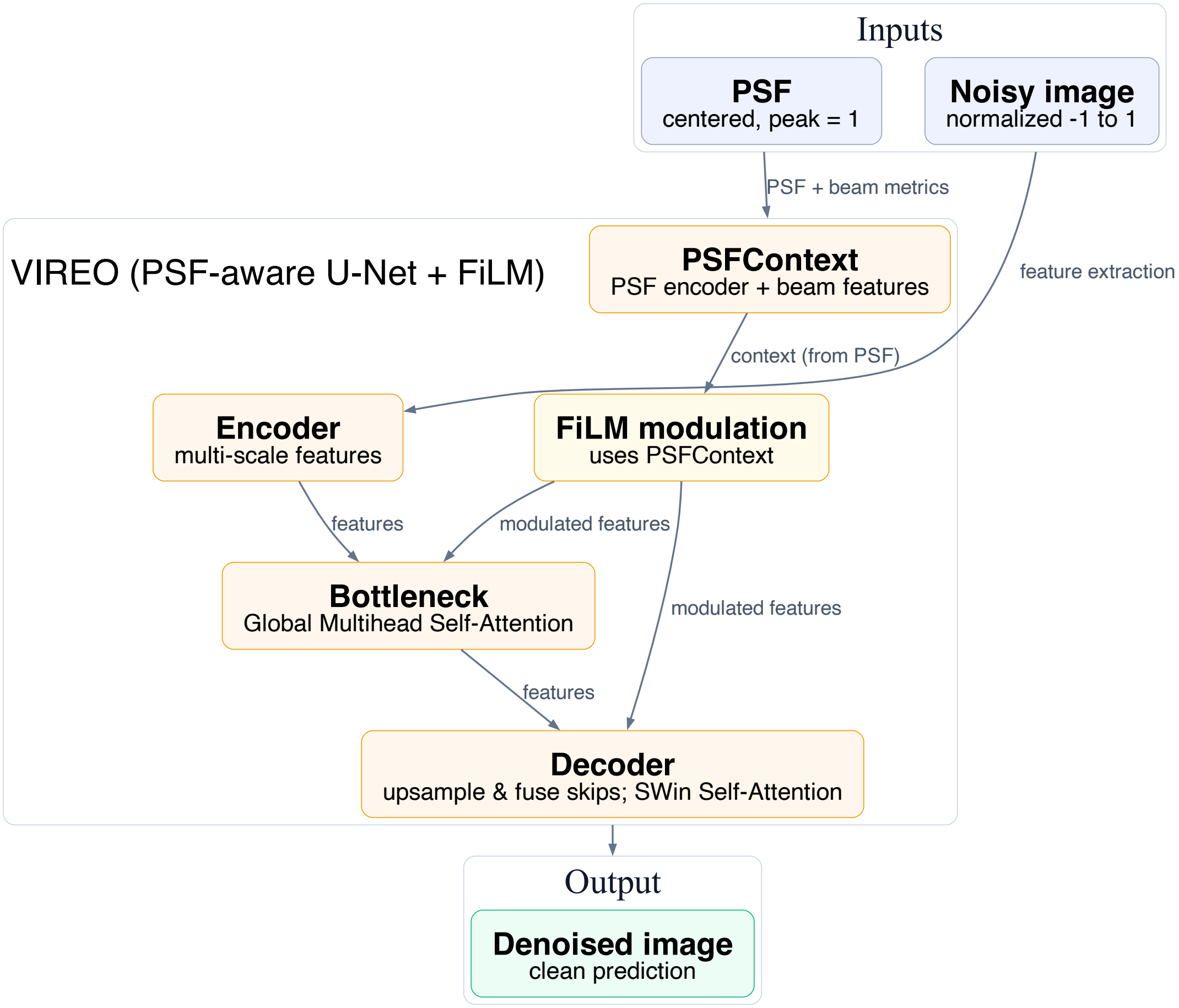

**Figure 7.** VIREO model general flow.

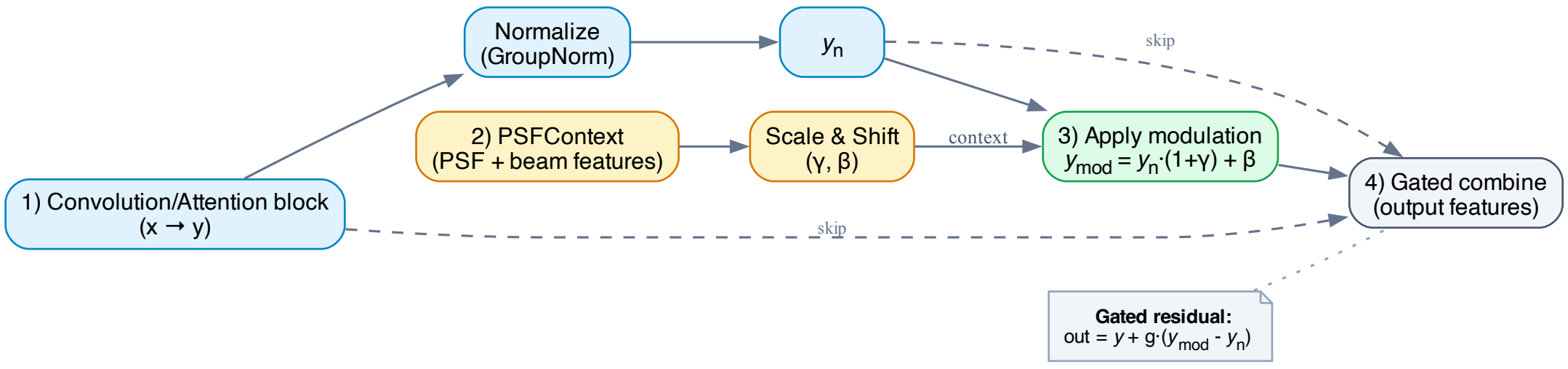

**Figure 8.** Overview of FiLM layer.

improvements relative to the noisy data, but VIREO always leads to the lowest RMS.

Denoised observations should also conserve the flux from the clean synthetic observation. This is measured in Figure 14 with summary diagnostics given in Table 2 and Table 3. Kernel density estimation of the fractional flux difference (left panel) shows that, whereas the noisy and CASA-cleaned data tend to underestimate the flux relative to the clean data, both the PSF-ignorant model and VIREO lead to outputs with fluxes that are close to the original synthetic observation. VIREO in particular leads to a sharply peaked distribution that is

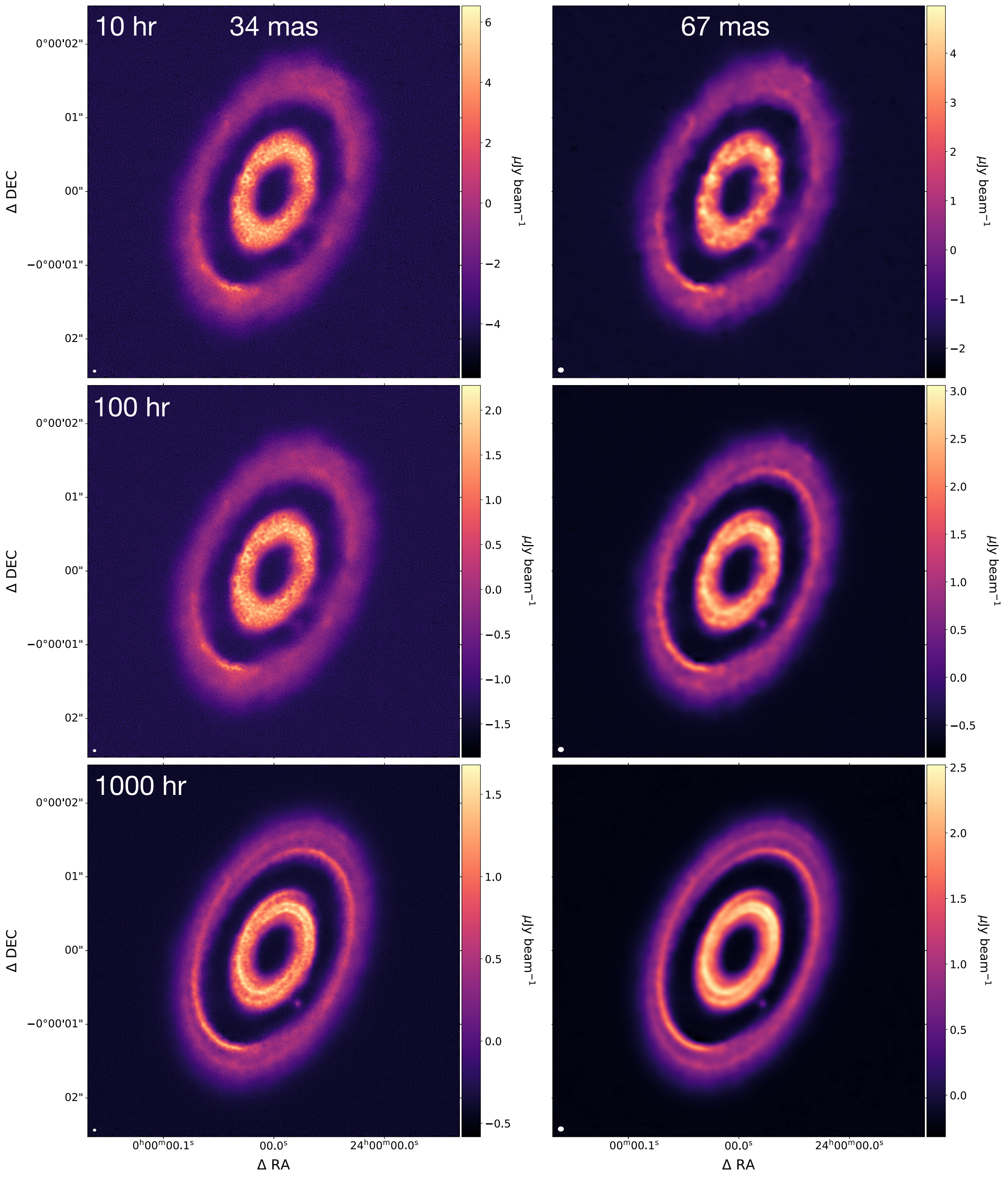

**Figure 9.** Denoised images corresponding to the clean image-plane observation in Figure 3 and the noisy images in Figure 4 and Figure 5. The left column is the 34 mas resolution data, and the right column is the 67 mas data. The top, middle, and bottom rows are 10 hour, 100 hour, and 1000 integration times, respectively. All images are significantly more clean than their noisy counterparts. Background noise is almost entirely removed, and substructures — rings, gaps, and spiral wakes — are made more prominent.

**Table 3.** Cumulative flux error metrics for each dataset.

| Dataset | Fraction with error ≤ 5% | Fraction with error ≤ 10% | 68$^{\mathrm{th}}$ error percentile | 95$^{\mathrm{th}}$ error percentile |
|---|---|---|---|---|
| Noisy data | 0.000 | 0.000 | 70.6% | 87.9% |
| VIREO cleaned | 0.769 | 0.935 | 3.7% | 11.9% |
| PSF-ignorant cleaned | 0.608 | 0.819 | 6.8% | 30.9% |
| `CASA` restored | 0.000 | 0.020 | 71.1% | 125.2% |

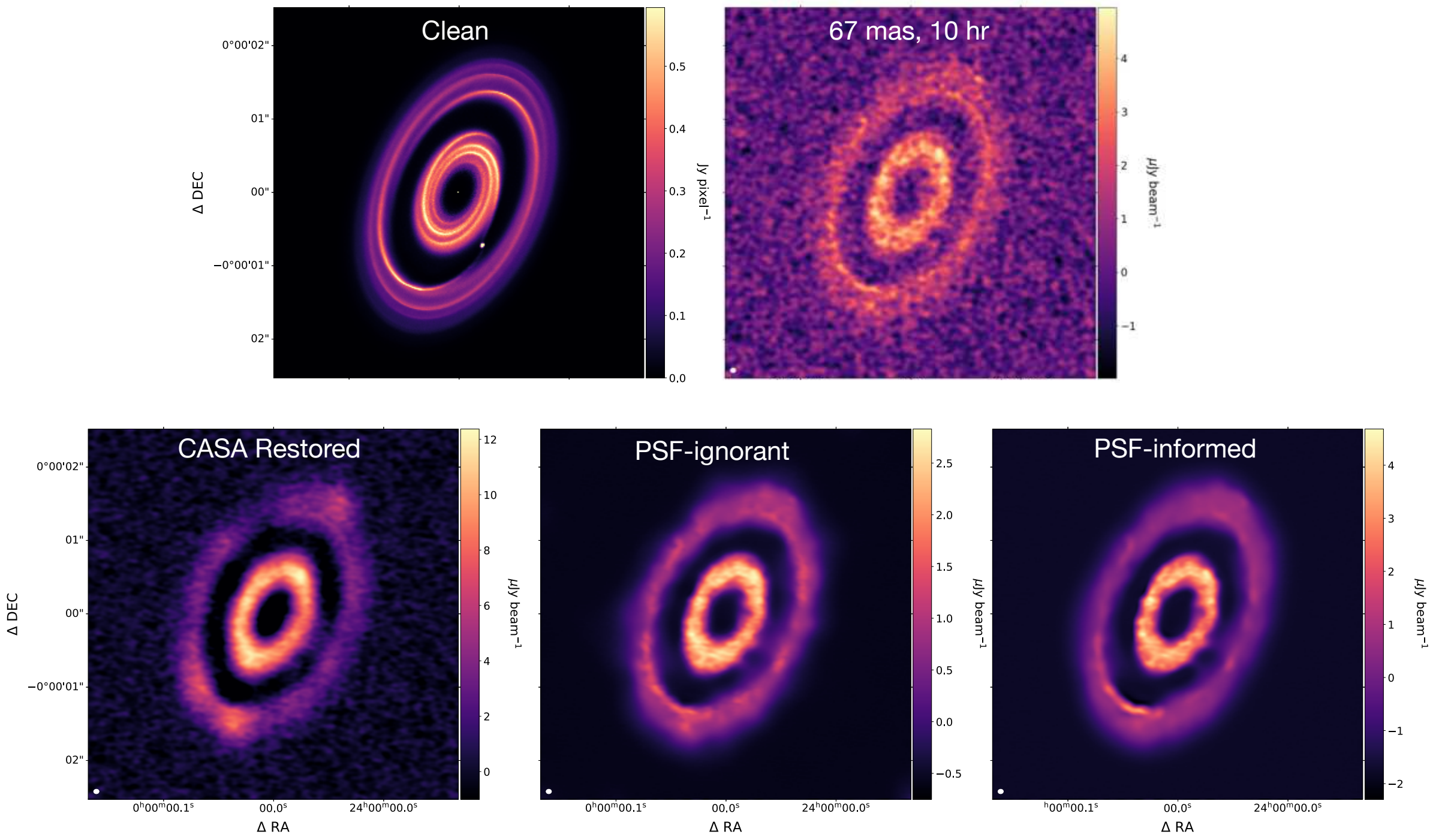

**Figure 10.** Comparison of CASA cleaning and PSF-informed denoising. Top left: clean image-plane target (same as Figure 3). Top right: 67 mas, 10 hour synthetic observation (same as second row in Figure 5) Bottom left: cleaned image using the CASA pipeline. Bottom middle: denoising output from PSF-ignorant model. Bottom right: VIREO output. While CASA significantly reduces noise and brings out substructure, VIREO removes background noise and more strongly brings out substructures like the planet's spiral wake. The PSF-ignorant model also removes noise, but it creates edge artifacts (e.g., at the inner edge of the outer ring) that can obscure substructure.

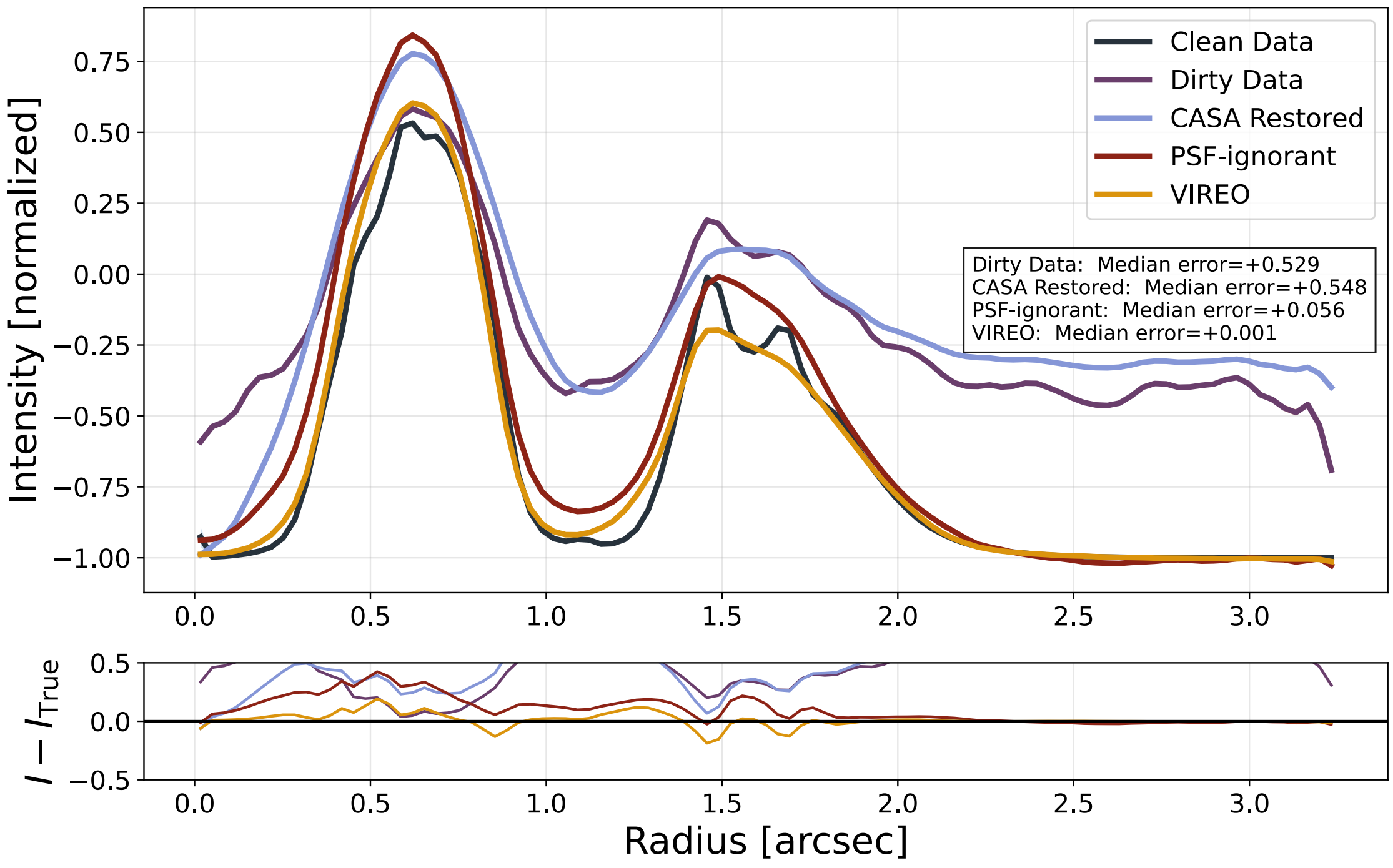

**Figure 11.** Azimuthally averaged radial profile from deprojected reconstructions shown in Figure 10. VIREO has the lowest median error and preserves the rings and gaps seen in the synthetic observation.

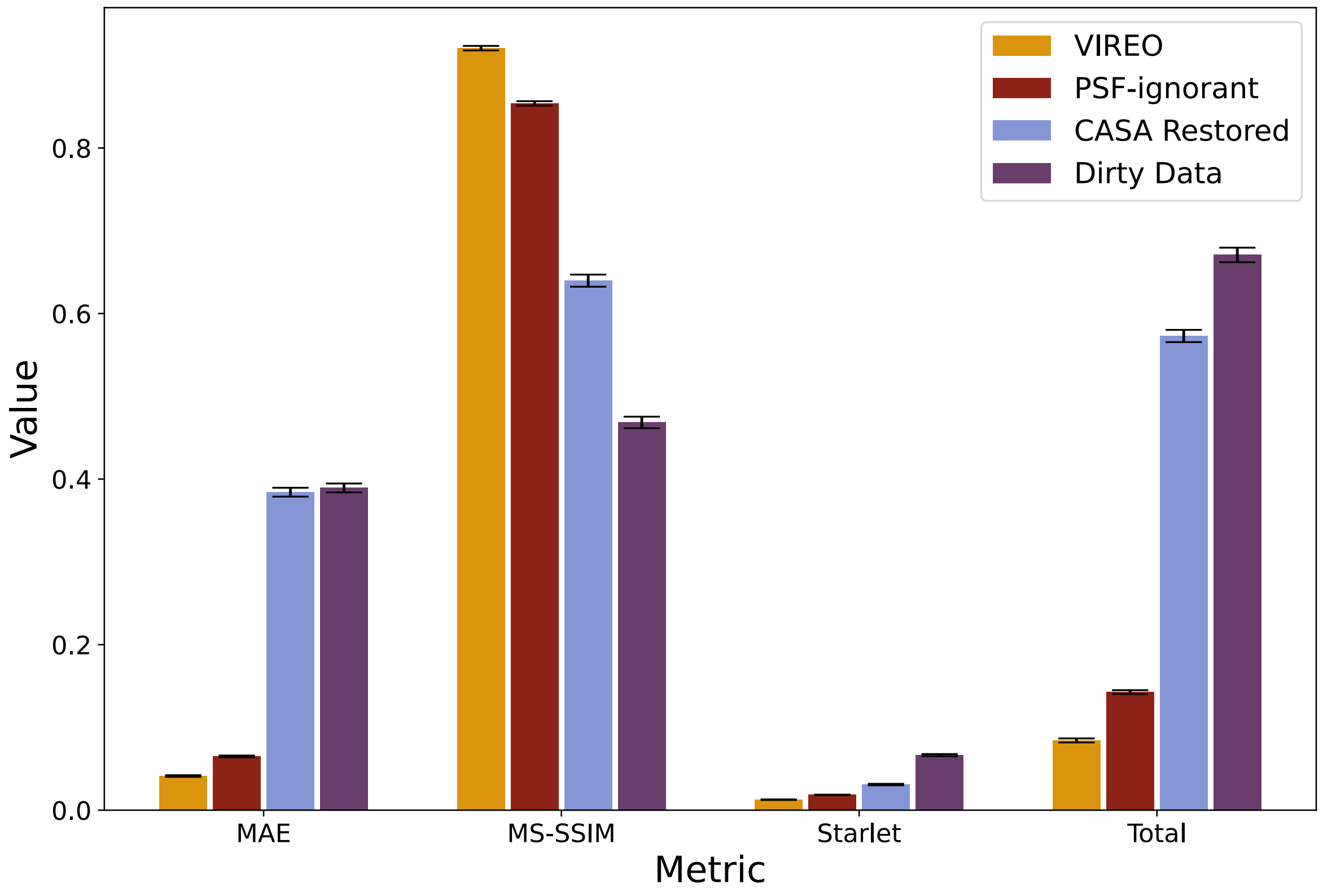

**Figure 12.** Visual metrics comparing the different datasets to the clean data. Total loss (right-most group) is defined as: $L_{total}$ = [MAE + 0.5(1 − MS_SSIM) + 0.25 × Starlet]. Orange: PSF-informed results. Red: PSF-ignorant model. Light blue: data deconvolved with CASA. Purple: original noisy data. Metrics were calculated through bootstrapping 80% of the test data 1,000 times. The height of the bar is the mean, and the error bars represent the standard deviation. In all cases, the PSF-informed model has the best metrics.

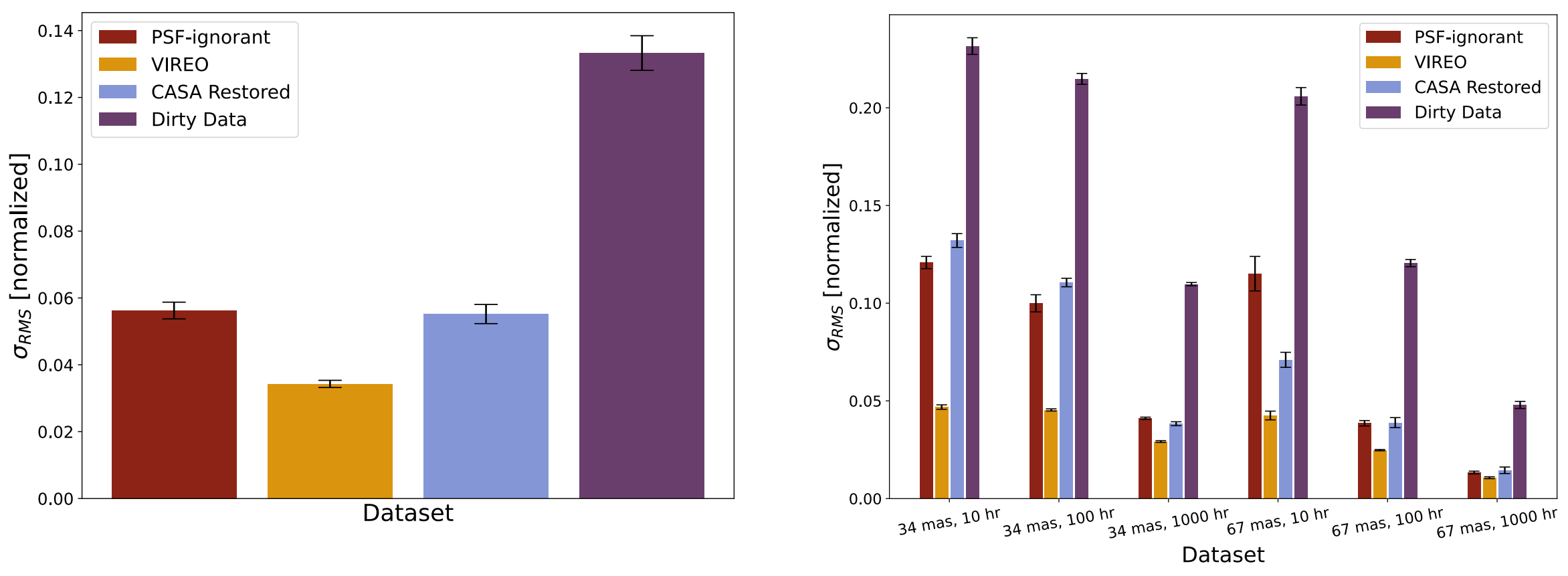

**Figure 13.** Corner RMS noise for the different datasets. Left: median RMS over all observations. Right: RMS by observational parameters. VIREO leads to both the lowest average RMS and the lowest RMS for all data subsets. Metrics are bootstrapped over 1,000 iterations over 80% of the test data, and the error bars represent 1 standard deviation.

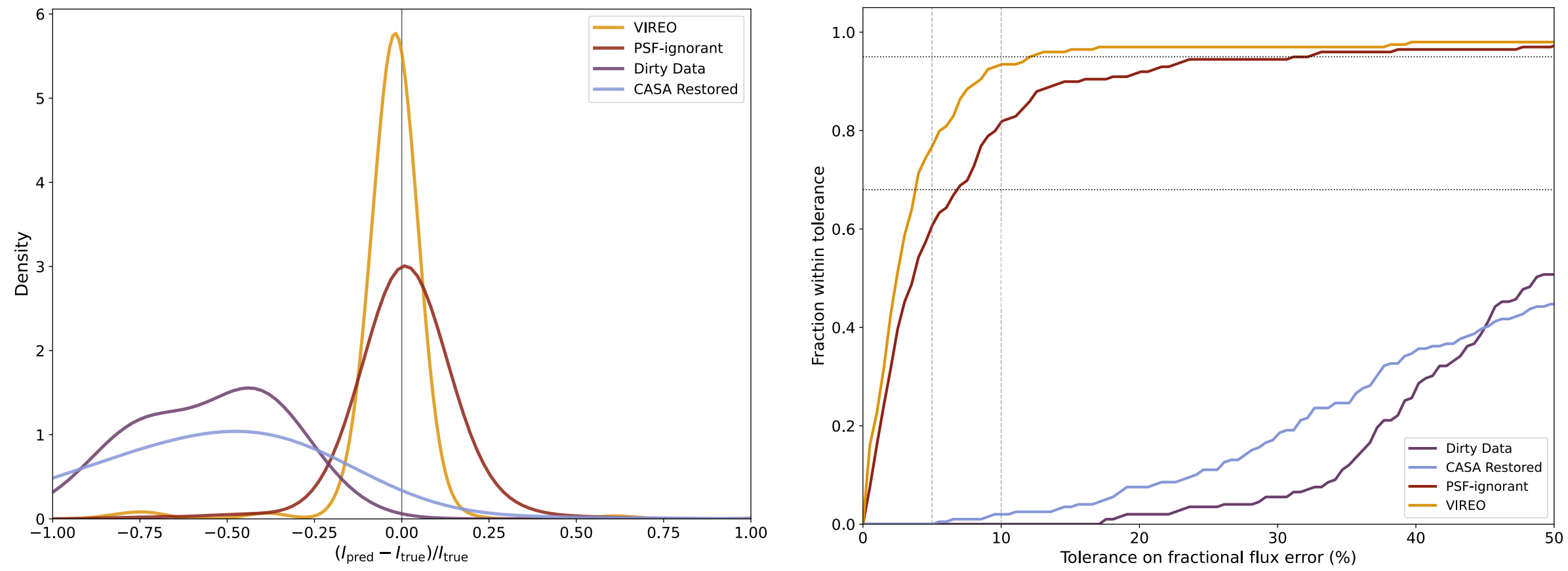

**Figure 14.** Flux-recovery metrics. Left: kernel density estimation plot of the fractional difference between the original image-plane observation flux and cleaned flux where the vertical line is at 0. Right: cumulative distribution of the absolute value of the fractional flux error. VIREO has a narrow error distribution that peaks at ≈0, and nearly 95% of the pixels have a flux error less than 10%.

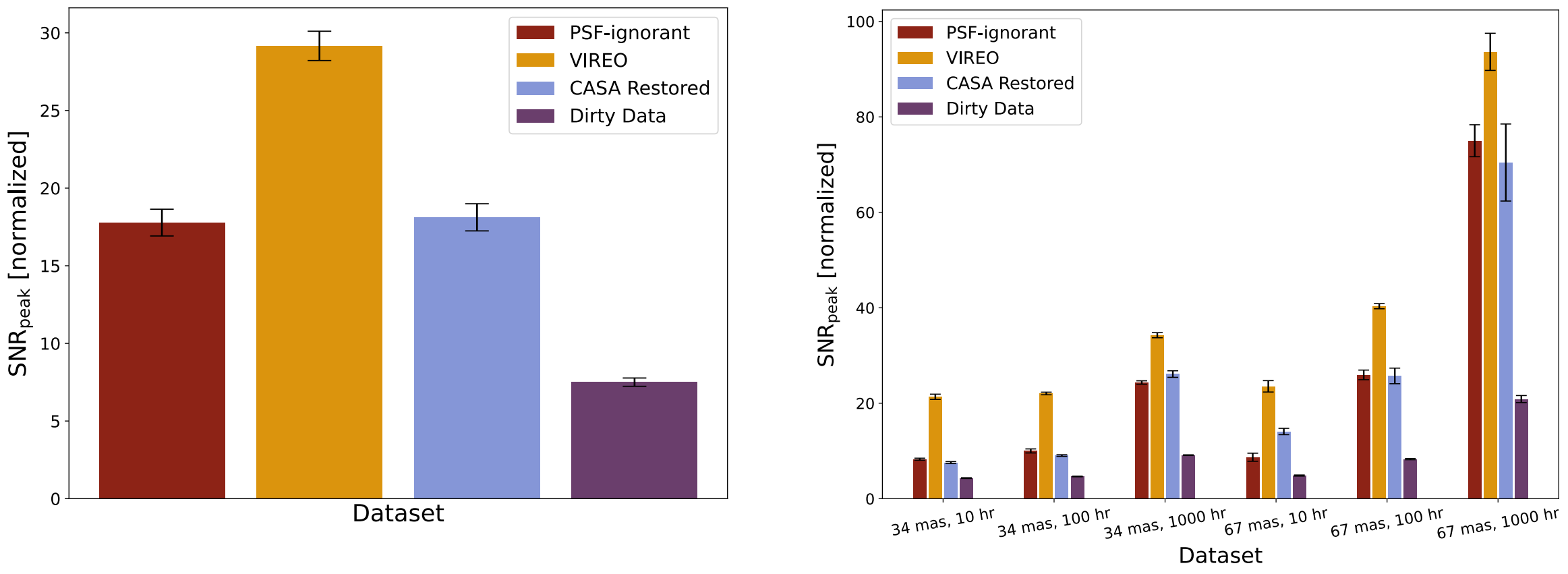

**Figure 15.** Peak signal-to-noise ratio for the different datasets. Left: median peak SNR over all observations. Right: peak SNR by observational parameters. VIREO has significantly larger peak SNR than all of our other denoising methods across all datasets. Metrics are bootstrapped over 1,000 iterations over 80% of the test data, and the error bars represent 1 standard deviation.

nearly centered at 0. In terms of both median and median absolute deviation (MAD), VIREO outperforms all of our other methods. The cumulative distribution of fractional flux error, i.e., the fraction of pixels that have an error less than some threshold, also shows that nearly 95% of VIREO output pixels have a flux error of less than 10%, larger than the corresponding fraction for the PSF-ignorant model.

A similar trend exists for the signal-to-noise ratio (SNR) of the denoised datasets. Figure 15 shows that VIREO has the highest peak SNR across all datasets. On average, the PSF-ignorant model has a higher peak SNR than the CASA-restored data, but CASA is higher for the cleanest datasets (67 mas, 100 and 1000 hours).

### 3.2 Analysis model results

To quantify the utility of our denoising methods not just in reconstruction metrics, but in the context of downstream disc analyses, we apply our classifier/regressor model to the cleaned outputs. Figure 16 shows these results. Other than the perfectly clean training data, VIREO's outputs are more useful for observational analysis, both in terms of classifying the disc as planet-hosting (AUC) and predicting the number of planets within the disc (MAE), compared to all other datasets. Figure 17 shows the confusion matrices for binary classification, and Figure 18 shows the regression confusion matrices. The regression confusion matrices indicate that the regressor is less conservative than the binary classifier in identifying planet-free discs. Predictions of $N_{\rm planet} = 0$ are relatively rare, so the regressor more often assigns non-zero counts to truly planet-free systems. This is likely because the classifier is explicitly optimized to distinguish $N_{\rm planet} = 0$ from $N_{\rm planet} > 0$, whereas the regressor is optimized over a continuous count space and can respond to weak residual substructure, noise, or reconstruction artifacts by predicting a small positive planet count. Thus, the count estimate should be interpreted as a conditional downstream diagnostic, best applied after the classifier has identified a disc as likely planet-hosting.

Both the binary and count-based confusion matrices show that the noisy, CASA-restored, and PSF-ignorant datasets are biased toward planet-hosting or higher-$N_{\rm planet}$ predictions. This improvement is consistent with the reconstruction metrics: VIREO has the lowest off-source RMS and highest peak SNR among the reconstructed datasets (Figures 13 and 15). By suppressing beam-correlated noise and residual artifacts while preserving coherent disc substructure, VIREO reduces the tendency of the downstream model to interpret noise features as planet-induced structures. We therefore consider the classifier/regressor results not as a production-ready planet-detection pipeline, but as evidence that PSF-informed reconstruction better preserves the image information used by downstream analysis models.

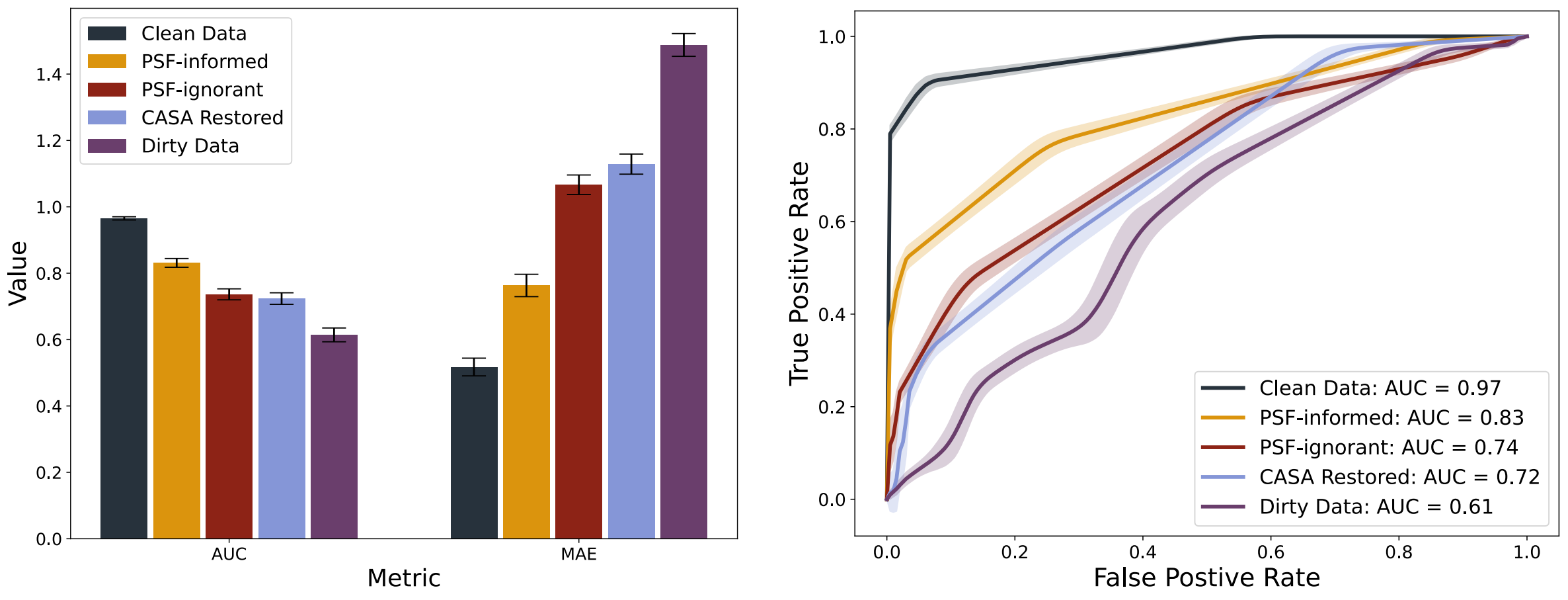

**Figure 16.** Performance metrics of the classification/regression model trained on the clean data when applied to the different datasets. Left: AUC and MAE. Right: ROC curves. Black: clean data. Orange: data denoised with PSF-informed model. Red: data denoised with PSF-ignorant model. Light blue: data deconvolved with `CASA`. Purple: original noisy data. Metrics were calculated by bootstrapping 80% of the test data 1,000 times. The height of the bar is the mean, and the error bars represent the standard deviation (similarly for the solid and shaded regions of the ROC curves). For these downstream metrics, VIREO gives the best performance among the reconstructed datasets considered here.

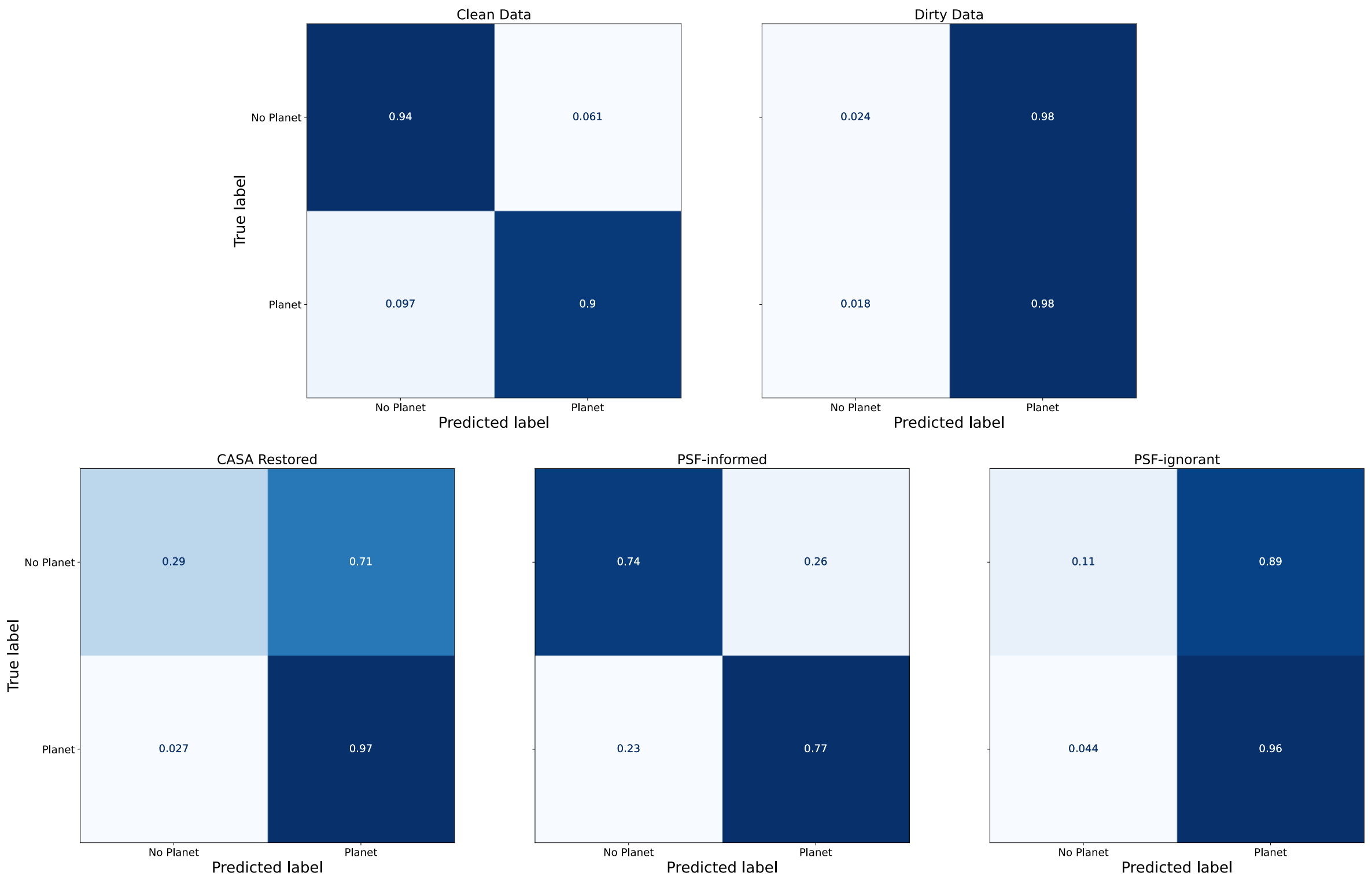

**Figure 17.** Confusion matrices of analysis model applied to all datasets where the values give the fraction of the data in a given box. Other than the clean training data, VIREO-reconstructed data shows the best balanced performance among the reconstructed datasets and substantially reduces the false-positive rate relative to the noisy, CASA-restored, and PSF-ignorant datasets.

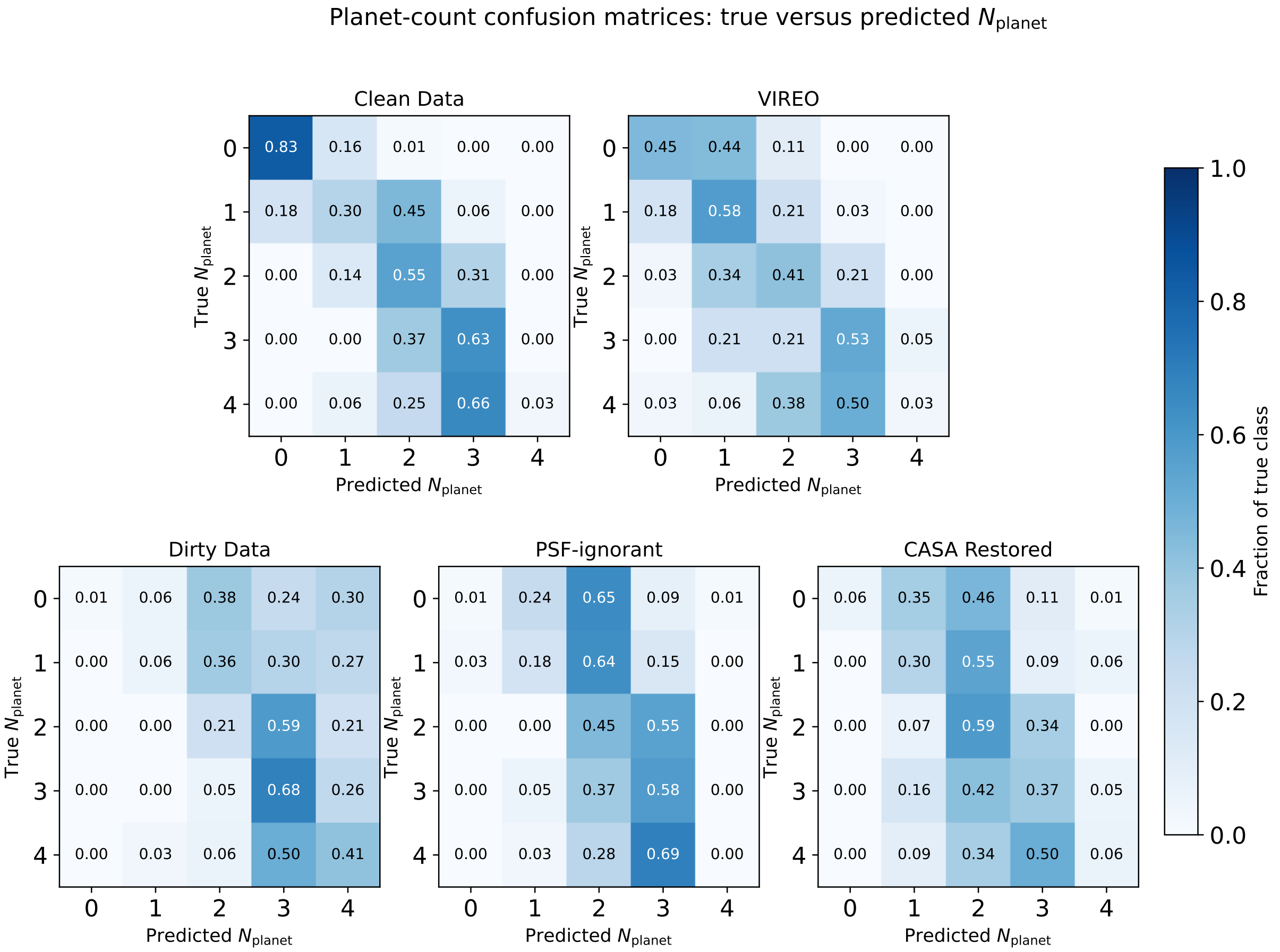

**Figure 18.** Confusion matrices of regression analysis model applied to all datasets where the values give the fraction of the data in a given box. VIREO most closely preserves the clean-data count structure among the reconstructed datasets, especially for the no-planet and low-multiplicity cases. Exact planet counting remains difficult for all inputs, including the clean images.

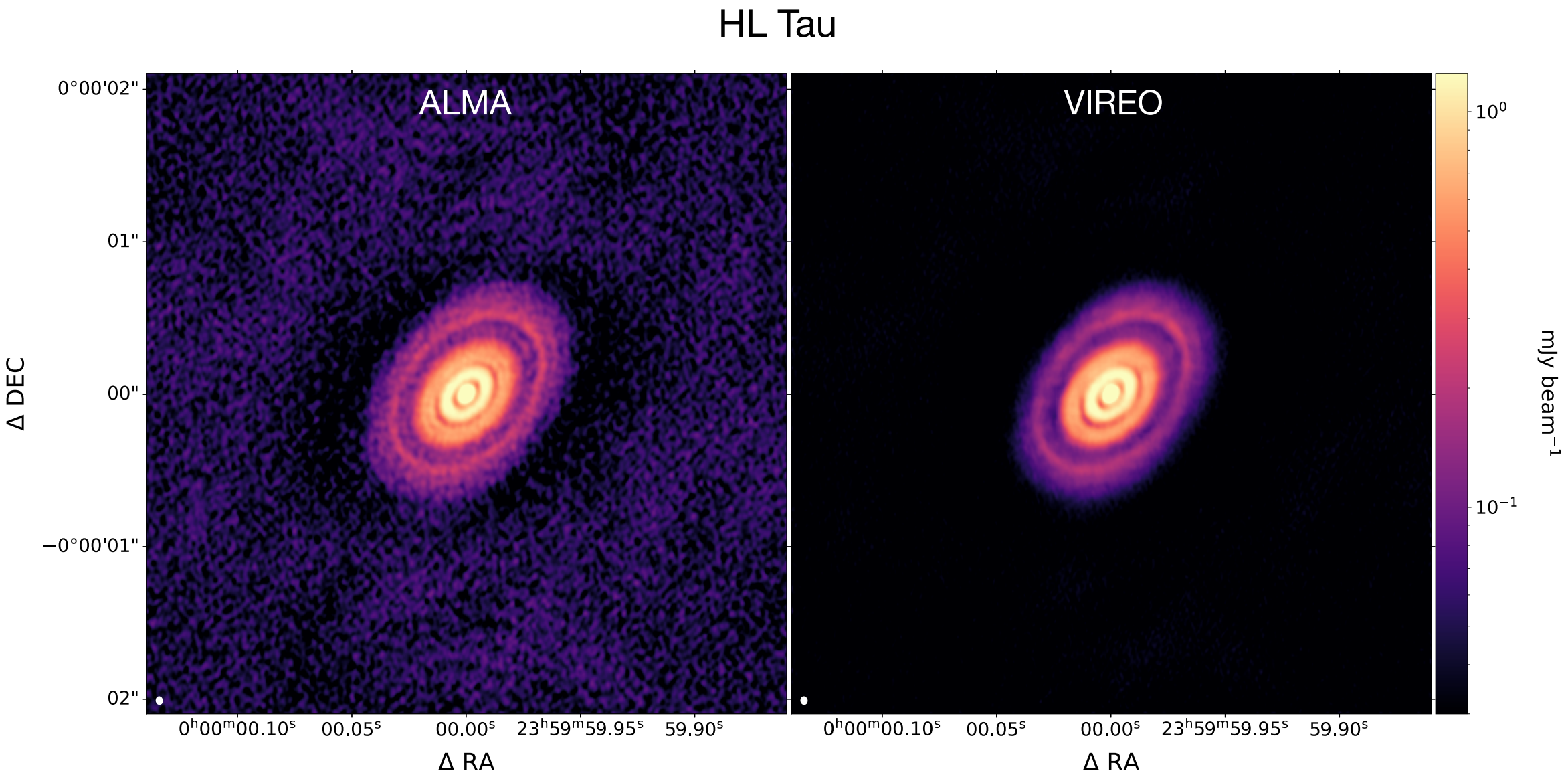

**Figure 19.** HL Tau denoising results. Left: ALMA data (ALMA Partnership et al. 2015b) Right: VIREO results. VIREO removes the background noise while preserving the annular substructure and the ring-gap contrast.

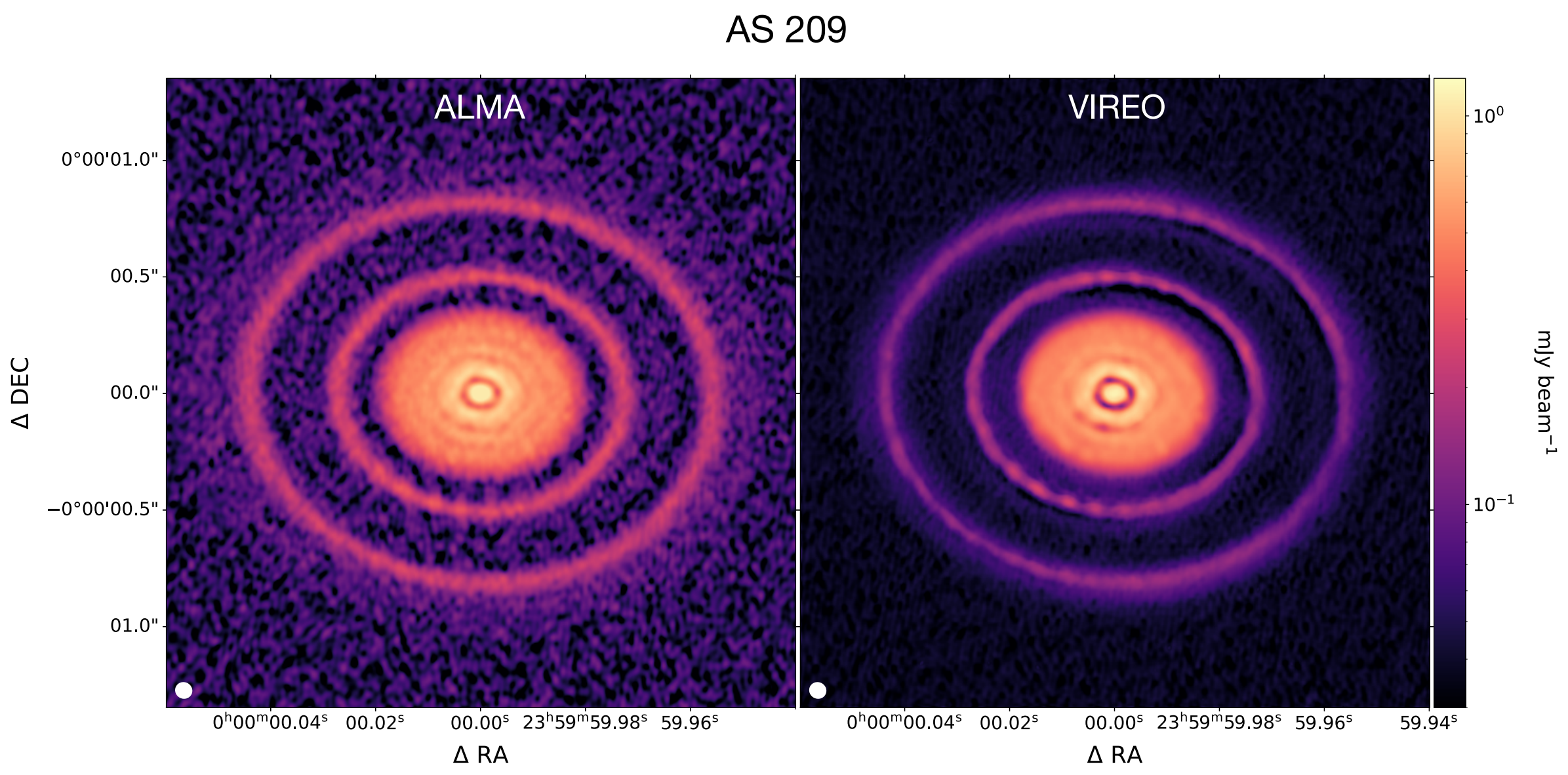

**Figure 20.** AS 209 denoising results. Left: ALMA data (Andrews et al. 2018) Right: VIREO results. Noise is reduced and the ring structure is preserved, with some ring peaks appearing sharper in the VIREO output.

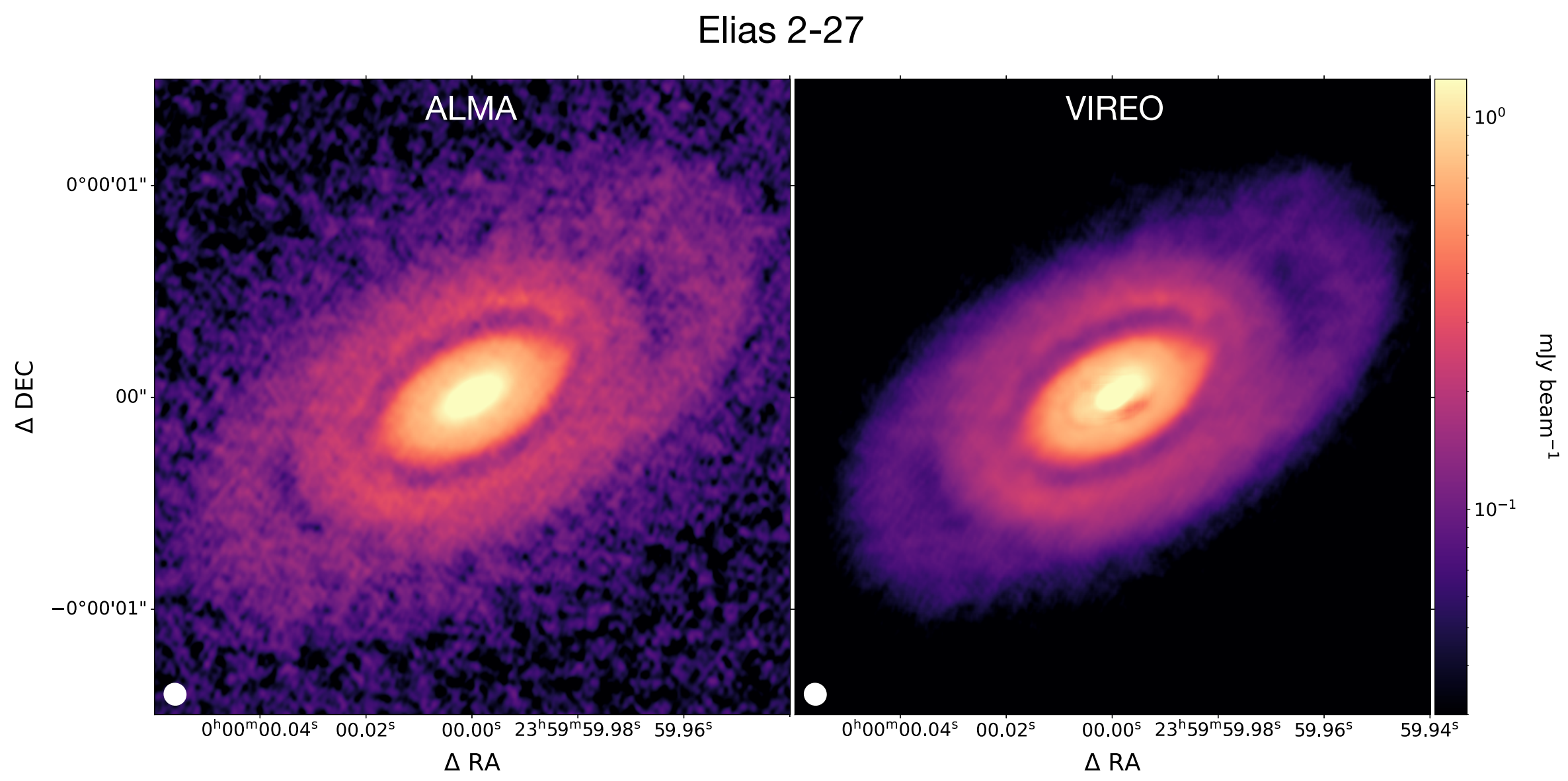

**Figure 21.** Elias 2-27 denoising results. Left: ALMA data (Andrews et al. 2018). Right: VIREO results. The VIREO output not only has less background noise, but it also increases the visibility of the spiral arms. The small darker feature near the middle of the disc may be a reconstruction artifact and should not be interpreted as a new feature.

### 3.3 Application to ALMA data

Since real SKAO-Mid observations do not yet exist, we test our model using the archival ALMA data. The archival ALMA images used here are restored image-plane science products rather than raw undeconvolved dirty maps. They therefore provide a real-data test in the same broad data-product regime targeted by our fiducial synthetic prescription, although they may contain calibration and deconvolution residuals, together with spatially varying noise and other effects associated with primary-beam correction.

We train the same model using identically prepared synthetic observations at a wavelength of 1.25 mm with a grain-size distribution of 0.03$\mu$m to 4 mm to approximate ALMA data. The overall method of generating our noisy ALMA data is the same as that for our SKAO-Mid data, and the simulations are unchanged, but we use the ALMA configuration and noise/SNR levels taken from the recent exoALMA data products (Teague et al. 2025). Using data prepared for ALMA is important because SKAO-Mid and ALMA observe different grain sizes and substructures at different scales. We found that applying a VIREO trained on SKAO-Mid data obscured some of the smaller substructures seen in ALMA data. Existing ALMA data is cleaner than most of the synthetic SKAO-Mid data we created, and, as shown in Section 3.1, the relative improvement of denoising diminishes as observations become cleaner. As such, we expect VIREO to lead to less of a qualitative improvement when applied to ALMA data when compared to the SKAO-Mid results. However, our results still show notable improvements.

In practice, observational data may already include the PSF. However, if this is not the case, we approximate a PSF by building an elliptical Gaussian using the beam major and minor axis provided by the observation's metadata. We compared the two methods using observations that provided the PSF and found negligible differences in the resulting image.

Figure 19, Figure 20, and Figure 21 show example outputs from VIREO applied to HL Tau (ALMA Partnership et al. 2015b), AS 209, and Elias 2-27 (Andrews et al. 2018), respectively. Further results for HD 142666 (Figure C1) and IM Lup (Figure C2) are given in Appendix C. While substructure is visible in all ALMA data, VIREO removes background noise while maintaining substructure and enhancing the contrast between the disc and the background. In HL Tau (Figure 19), VIREO decreases the background noise while preserving annular substructure. When applied to AS 209 (Figure 20), VIREO sharpens some ring peaks and reduces background noise. VIREO increases the apparent visibility of the spiral arms in Elias 2-27 (Figure 21) and generally makes the outer disc stand out more strongly against the background. As expected, the high quality of ALMA outputs leads to VIREO acting as more of a noise suppressor that preserves substructure rather than a method of qualitatively improving the observation for downstream analysis. Still, VIREO's performance on ALMA data is promising for its deployment on noisier SKAO data. We note that some of the VIREO-reconstructed images, e.g. Elias 2-27 (Figure 21), show small reconstruction artifacts near the middle of the disc. While they are generally small (approximately the size of the beam) and noticeable, in practice, one should be careful to not interpret them as true disc features.

Deprojected azimuthally average radial intensity profiles for AS 209 (Figure 20) and HL Tau (Figure 19) are shown in Figure 22. In both cases, VIREO preserves the flux and contrast by removing background noise. For example, when a Gaussian profile is fit to the ring in AS 209 at ≈0.6 arcsec, the standard deviation for the ALMA profile is over 18% larger than the fit of the VIREO profile, showing a non-negligible increase in the sharpness of the ring. Figure 23 shows that VIREO additionally decreases RMS relative to the ALMA input.

## 4 DISCUSSION

Our results suggest that visibility-informed models reconstruct clean observations better than the traditional methods and the denoising machine learning-based models that are ignorant of the observation's PSF. A U-Net with FiLM layers and PSF-blur loss produces denoised observations that preserve substructure, recreate average brightness, and match clean synthetic observation according to both typical image similarity metrics and performance in a classifier/regressor model.

Even though a PSF-ignorant model still offers some improvement over traditional cleaning (Figure 12 and Figure 16), introducing the PSF through FiLM layers and convolution loss leads to improvement in all metrics. PSF-ignorant models have worse image similarity measures and do not perform as well when used to analyse the discs. Cleaning using `CASA` improves image clarity and utility, but VIREO gives the best overall performance for the metrics considered. The PSF-ignorant U-Net offers some improvement over `CASA` in terms of loss metrics, but both methods lead to less improvement than VIREO. Furthermore, denoising a single observation with VIREO is nearly 800 times faster than a `CASA` run (≈ 0.07 seconds versus ≈ 59 seconds when averaged over 100 samples).

While VIREO improves all the images, it is most valuable for the lowest-quality observations (e.g., 10 hour observations with 34 mas resolution). Figure 4, Figure 5, Figure 14, Figure 15, and the plots in Appendix B show this. Fine substructures that are nearly invisible in the noisiest observations come out clearly when passed through VIREO. Conversely, observations that already have clear substructure and little noise (e.g., 1,000 hour observations with 67 mas resolution) have less room for qualitative improvement, and VIREO results more closely resemble the original inputs with less background noise and sharpened substructure. However, this is encouraging because it implies that robust data analysis is possible without resorting to observations spanning several weeks; realistic and practical observation times can still create quality, usable data products when cleaned with VIREO. Using VIREO will allow for more observations, either of the same target or different objects entirely, without sacrificing analysis. This would significantly increase the efficiency of SKAO-Mid and increase its data output.

While this work used training data designed to replicate SKAO-Mid observations, applying VIREO to archival ALMA data demonstrates that it is capable of cleaning real observations, approximately preserving their measured radial flux profiles and resolved substructure. Since our results suggest that VIREO leads to greater relative improvement as images become noisier, we expect that its application to SKAO-Mid observations will generate data with even greater qualitative enhancement.

A limitation of the present synthetic observations is that they are not full end-to-end visibility-domain simulations. The input images approximate restored image-plane products by convolving the clean radiative-transfer model with a Gaussian restoring beam and adding beam-correlated residual noise scaled to the adopted RMS and peak-SNR values. This captures the image-domain resolution and noise level relevant for the denoising task considered here, but it does not fully capture all features of real interferometric imaging. In particular, the fiducial synthetic data do not explicitly model residual dirty-beam sidelobes, imperfect deconvolution artifacts, calibration residuals, or the spatially varying noise introduced by primary-beam correction. These effects may be important in some observations. The applications to archival ALMA images provide an empirical test on real restored image products. Future work will use full measurement-set simulations and imaging pipelines to quantify how these additional residual structures affect both absolute reconstruction performance and the relative benefit of PSF conditioning.

Despite this limitation, the controlled comparison between the PSF-informed and PSF-ignorant denoising models remains a valid test of whether PSF conditioning provides useful information within the adopted restored-image distribution. Both models are trained and evaluated on the same synthetic observations and clean image-plane targets; the key methodological difference is that VIREO receives the UV-derived PSF through the FiLM conditioning and PSF-based loss term, whereas the comparison U-Net does not. Therefore, while the fiducial synthetic observations are an idealized restored-image approximation, the improved performance of VIREO over the PSF-ignorant model demonstrates that information about the interferometric response is useful for image-domain denoising within this data-product regime. The present comparison does not establish that the magnitude of this advantage will remain unchanged for every residual-artifact distribution, which will require end-to-end simulated observations.

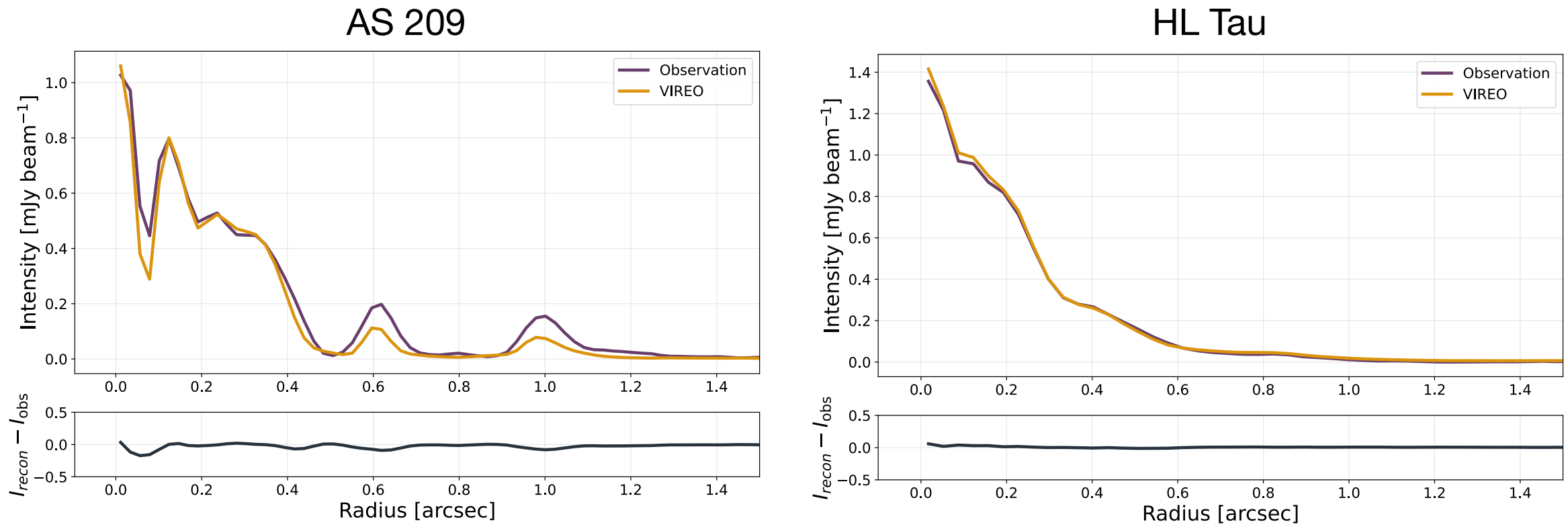

**Figure 22.** Deprojected azimuthally averaged radial intensity profiles. Left: AS 209. Right: HL Tau. VIREO preserves substructure visibility while maintaining flux and reducing noise. In particular, the peaks in AS 209 are narrower in the VIREO output and the background noise is reduced. The bottom of each panel shows the difference between the reconstructed and observed profiles.

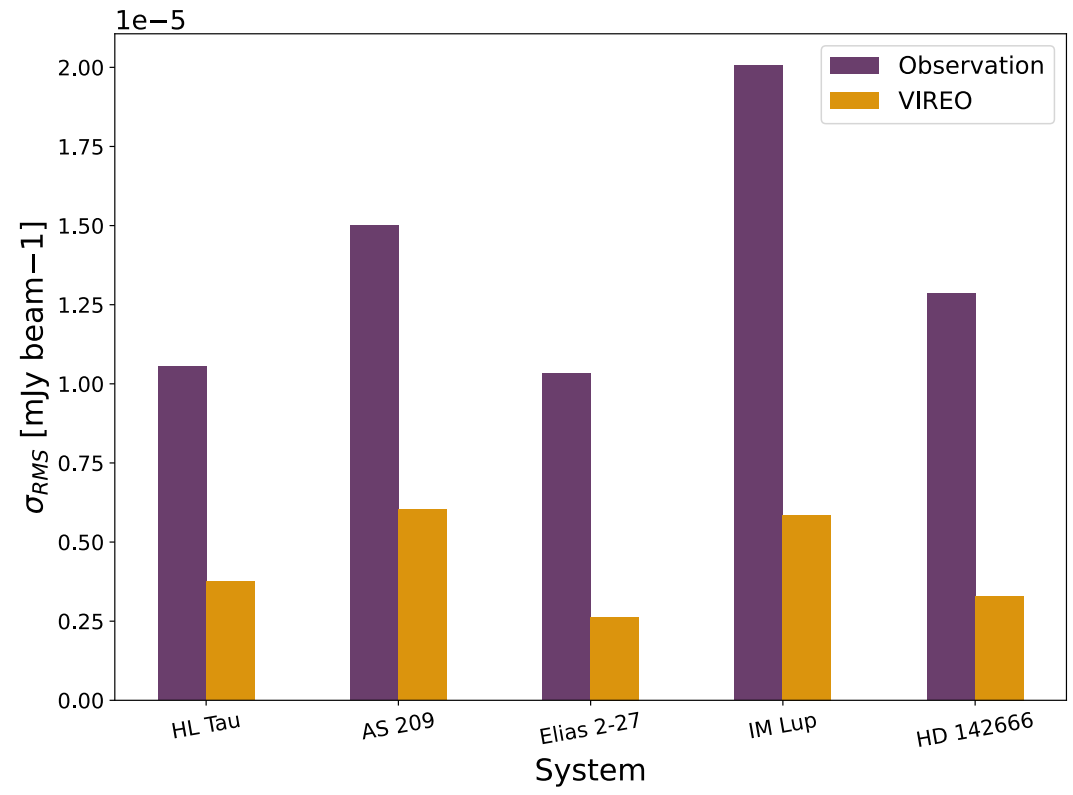

**Figure 23.** RMS noise for the observed and denoised systems. VIREO significantly decreases RMS noise across all systems.

## 5 CONCLUSIONS

We have demonstrated that the visibility-informed reconstructions of the interferometric observations are both more faithful reconstructions of perfectly clean data and improve the sensitivity and performance of downstream analyses of disc properties compared to traditional cleaning methods. Including the observation's PSF in the denoising pipeline improves the performance relative to PSF-ignorant methods. By the loss metrics, the observational metrics, the utility in analysis models, and the visual inspection, the visibility-informed model outperforms all the other reference methods used in this work.

Since SKAO-Mid is anticipated to begin observations in the next few years and previous works have shown that a robust analysis of protoplanetary discs with its data is difficult and complex (Ilee et al. 2020), developing the methods to improve data quality and analytical capabilities is paramount. Machine learning offers a way forwards that outperforms traditional methods, and its performance is maximized when information about the observation itself is included. Given the generality of this approach for all interferometric observations and its performance on ALMA data, our results suggest that the visibility-informed networks are applicable as a general technique for the cleaning and data analysis across facilities when trained on the appropriate datasets.

## ACKNOWLEDGMENTS

The authors would like to thank the anonymous reviewer for their thorough and helpful comments. This paper makes use of the following ALMA data: ADS/JAO.ALMA#2011.0.00015.SV. and ADS/JAO.ALMA#2016.1.00826.S. ALMA is a partnership of ESO (representing its member states), NSF (USA) and NINS (Japan), together with NRC (Canada), MOST and ASIAA (Taiwan), and KASI (Republic of Korea), in cooperation with the Republic of Chile. The Joint ALMA Observatory is operated by ESO, AUI/NRAO and NAOJ. The National Radio Astronomy Observatory is a facility of the National Science Foundation operated under cooperative agreement by Associated Universities, Inc. This study was supported in part by resources and technical expertise from the Georgia Advanced Computing Resource Center, a partnership between the University of Georgia's Office of the Vice President for Research and Office of the Vice President for Information Technology. Simulation results were visualized with `ProtoMatics` Terry (2023).

## DATA AVAILABILITY

The VIREO code is available at this repository: https://github.com/j-p-terry/vireo for public use and download (with proper citation under the BSD-3-Clause license). This includes the scripts used to generate the training data from radiative transfer outputs. The simulational data used in this article will be shared on reasonable request to the corresponding author.

## APPENDIX A: CASA SETTINGS

Table A1 gives information on the exact parameters used in the CASA pipeline used in this work.

## APPENDIX B: EXAMPLE SYNTHETIC OBSERVATION RESULTS

Here we present other example results from VIREO applied to synthetic observations.

## APPENDIX C: EXAMPLE ALMA OBSERVATION RESULTS

Here we present other example results from VIREO applied to archival ALMA observations.

## APPENDIX D: DETAILED VIREO ARCHITECTURE

Here we present a detailed view of the VIREO architecture.

**Table A1.** CASA multiscale deconvolution settings for the baseline pipeline, including the automatic second-pass depending on resulting RMS.

| Stage | Parameter | Value / Setting |
|---|---|---|
| Inputs | Residual HDU / PSF HDU | 0 / 1 (PSF max > 0 sanity check) |
| Units | Working unit | Jy beam$^{-1}$ (convert from SI by $\times 10^{-26}$) |
| | Export unit | W m$^{-2}$ Hz$^{-1}$ beam$^{-1}$ (post-processing) |
| Pixel grid | Cell size | From WCS CDELT, arcsec pix$^{-1}$; warn if $<$ 3 px/beam |
| PSF / Beam | Restoring beam | Gaussian fit to PSF core (BMAJ, BMIN, BPA); fallback header |
| Noise $\sigma$ | Region | Four corner boxes, width = 5% of $\min(N_x, N_y)$ |
| | Estimator | Sigma-clipped RMS (Jy beam$^{-1}$) |
| Mask | Seed | SNR > 4.0 on SNR map ($I/\sigma$) |
| | Grow kernel | Gaussian, 0.5 × beam (major, minor) |
| | Re-threshold | Keep pixels > 0.3; saved as `.mask` |
| | Use in CLEAN | `CASA` auto-detects `.mask` |
| Deconvolver | Algorithm | `deconvolve` with `deconvolver=multiscale` |
| | Scales (pixels) | $\{0, \lfloor b/4 \rfloor, \lfloor b/2 \rfloor, \lfloor b \rfloor, 2b, 4b, 8b\}$; integers, unique, capped at $0.35 \times \min(N_x, N_y)$ |
| | Iterations (pass 1) | $\leq$ 20,000 (or $n_{\rm iter}/3$) |
| | Gain (pass 1) | 0.05 |
| | Threshold (pass 1) | 1.2 $\sigma$ Jy |
| Second pass | Use | Only if decision rule is met |
| | Decision rule | Run pass 2 iff $\max(I_{\rm resid}\ \text{in mask})/\sigma_2 \geq 3.5$ |
| | Mask rebuild | Optional: SNR > 1.2 × 4.0 before pass 2 |
| | Scales (pixels) | $\{0, \lfloor b/4 \rfloor, \lfloor b/2 \rfloor, \lfloor b \rfloor, 2b, 4b\}$ |
| | Iterations (pass 2) | $\leq$ 25,000 |
| | Gain (pass 2) | 0.03 (gentler) |
| | Threshold (pass 2) | $\min(1.2\,\sigma_2,\ 1.2\,\sigma)$ Jy |
| Restoration | Formula | Restored = (model ⊗ restoring beam) + residual |
| Cosmetic | Post-smooth | 1.00 × beam FWHM |
| Version | `CASA` | 6.7.0.31 |

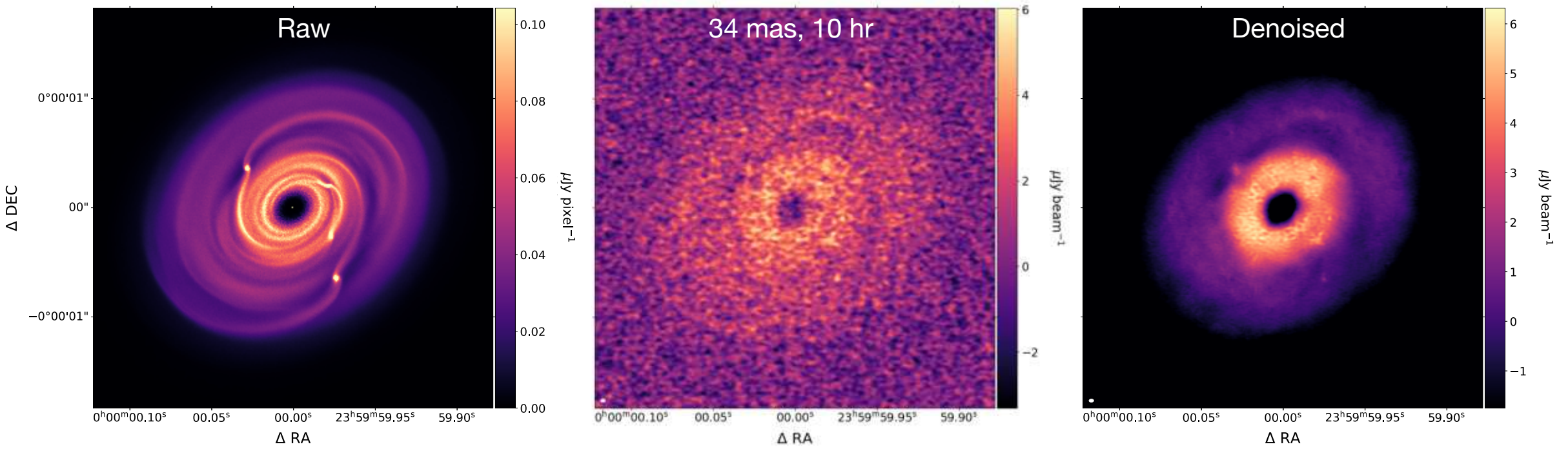

**Figure B1.** Denoising results for a synthetic observation with 34 mas resolution and 10 hours of total exposure time. Left: raw image-plane observation. Center: noisy observation. Right: VIREO results.

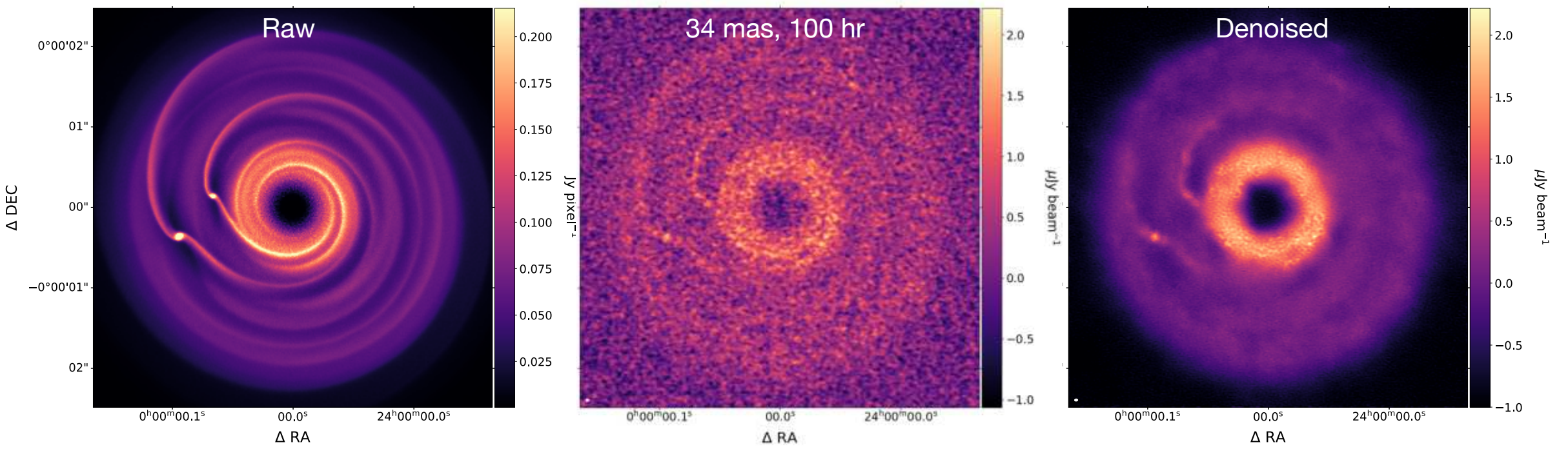

**Figure B2.** Same as Figure B1 except that the exposure time is 100 hours.

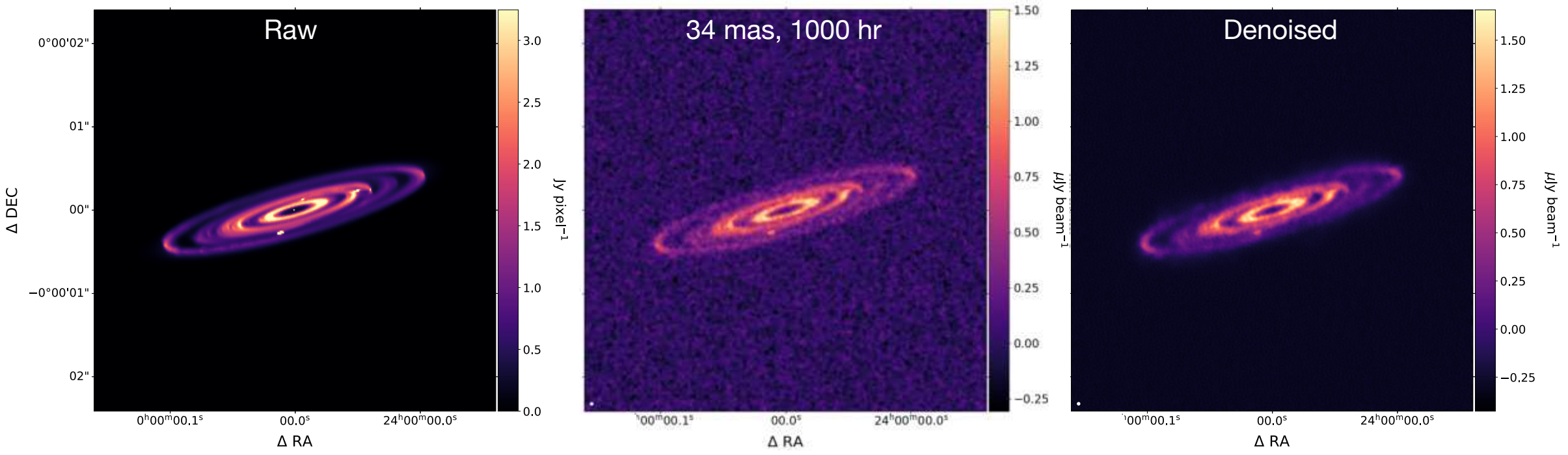

**Figure B3.** Same as Figure B1 except that the exposure time is 1000 hours.

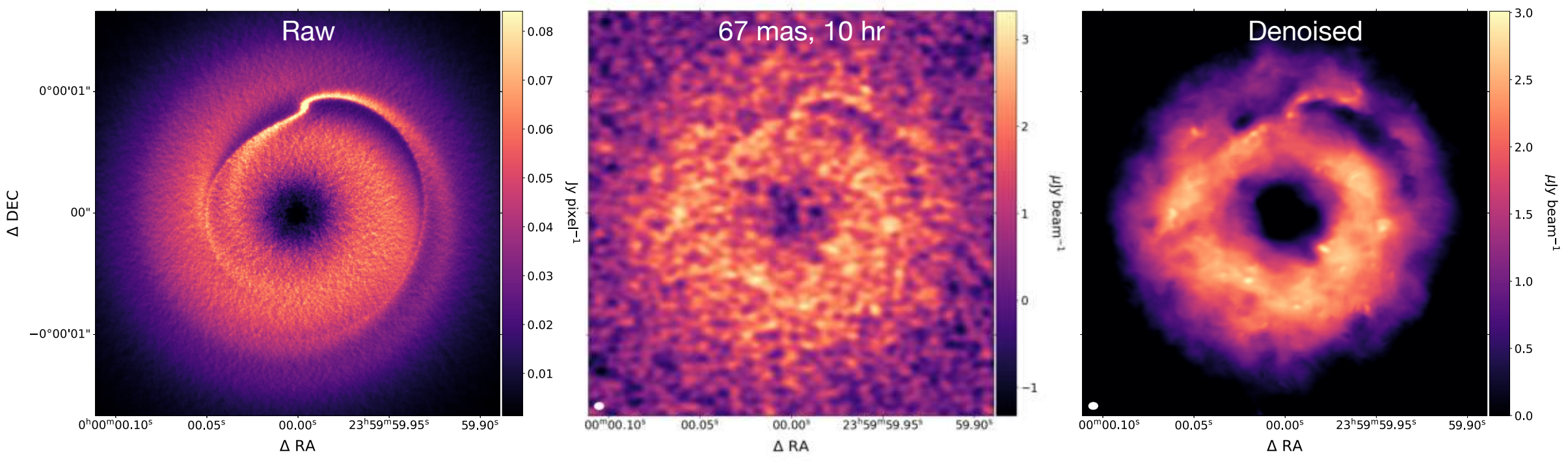

**Figure B4.** Denoising results for a synthetic observation with 67 mas resolution and 10 hours of total exposure time. Left: raw image-plane observation. Center: noisy observation. Right: VIREO results.

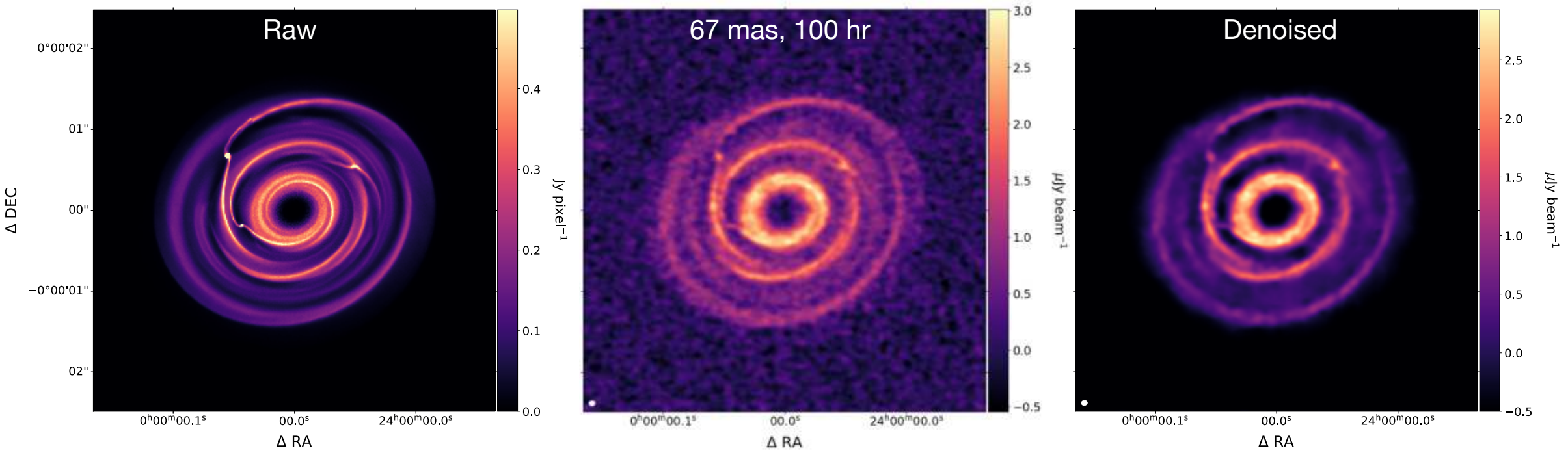

**Figure B5.** Same as Figure B4 except that the exposure time is 100 hours.

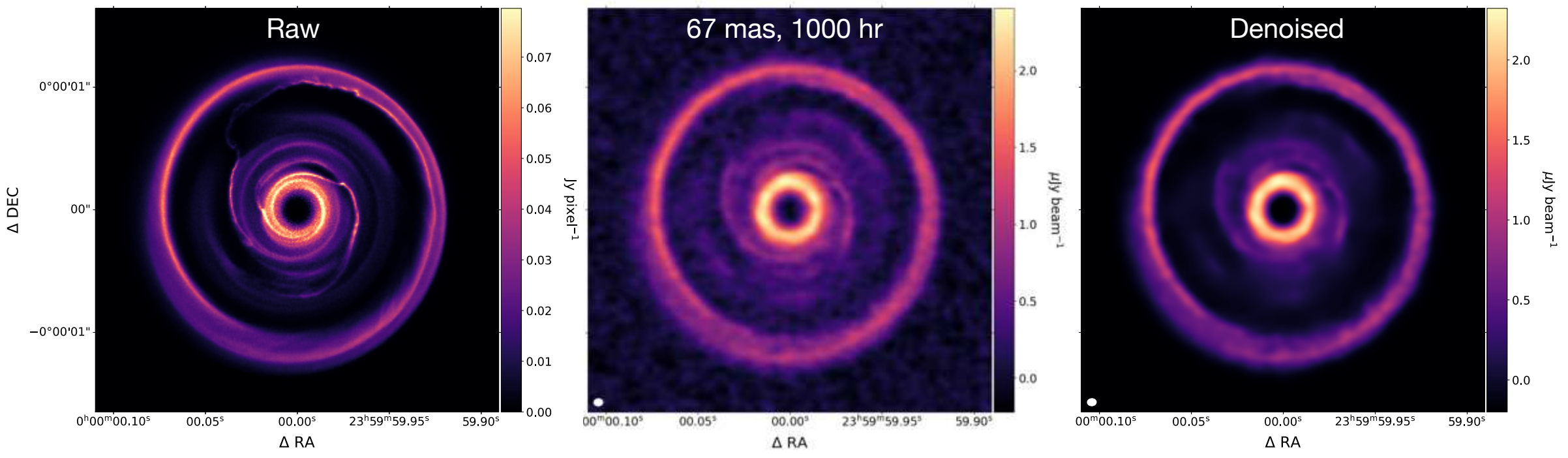

**Figure B6.** Same as Figure B4 except that the exposure time is 1000 hours.

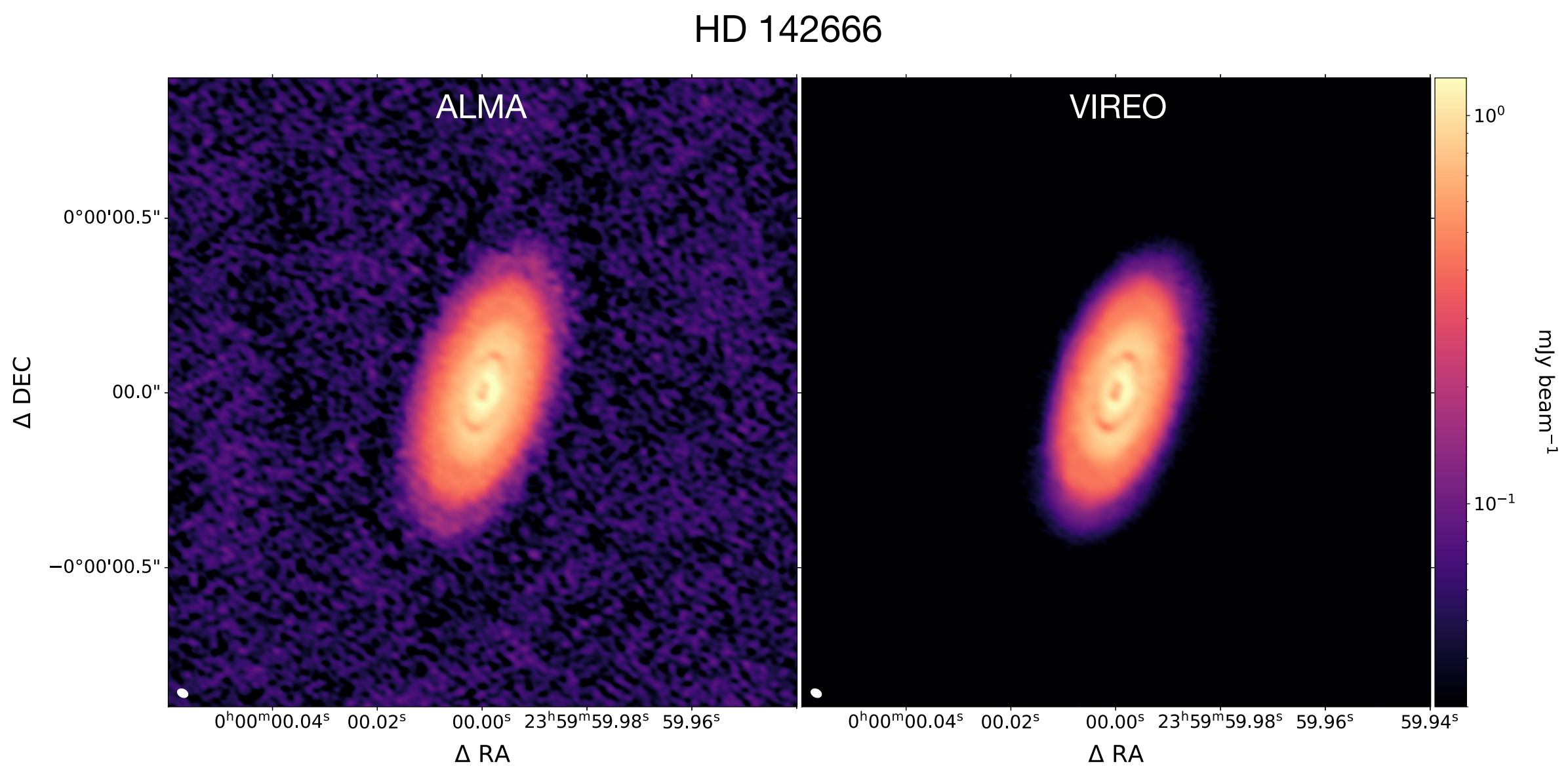

**Figure C1.** HD 142666 denoising results. Left: ALMA data (Andrews et al. 2018). Right: VIREO results. Background noise is almost completely removed while substructure and flux are both preserved.

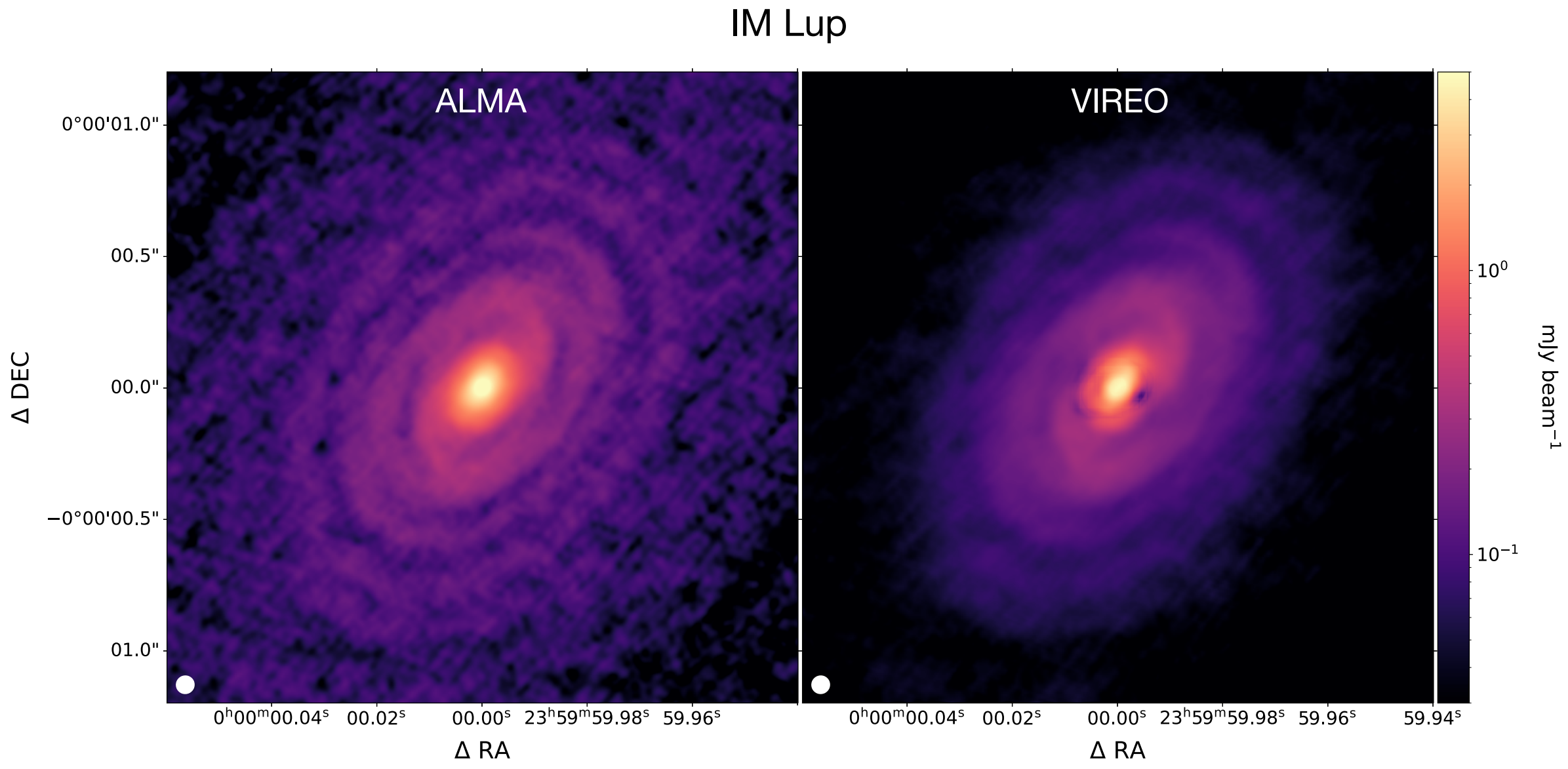

**Figure C2.** IM Lup denoising results. Left: ALMA data (Andrews et al. 2018). Right: VIREO results. VIREO removes noise and enhances the spiral substructure. The small black feature near the middle of the disc is likely a reconstruction artifact, and we do not consider this to be a true feature of the disc.

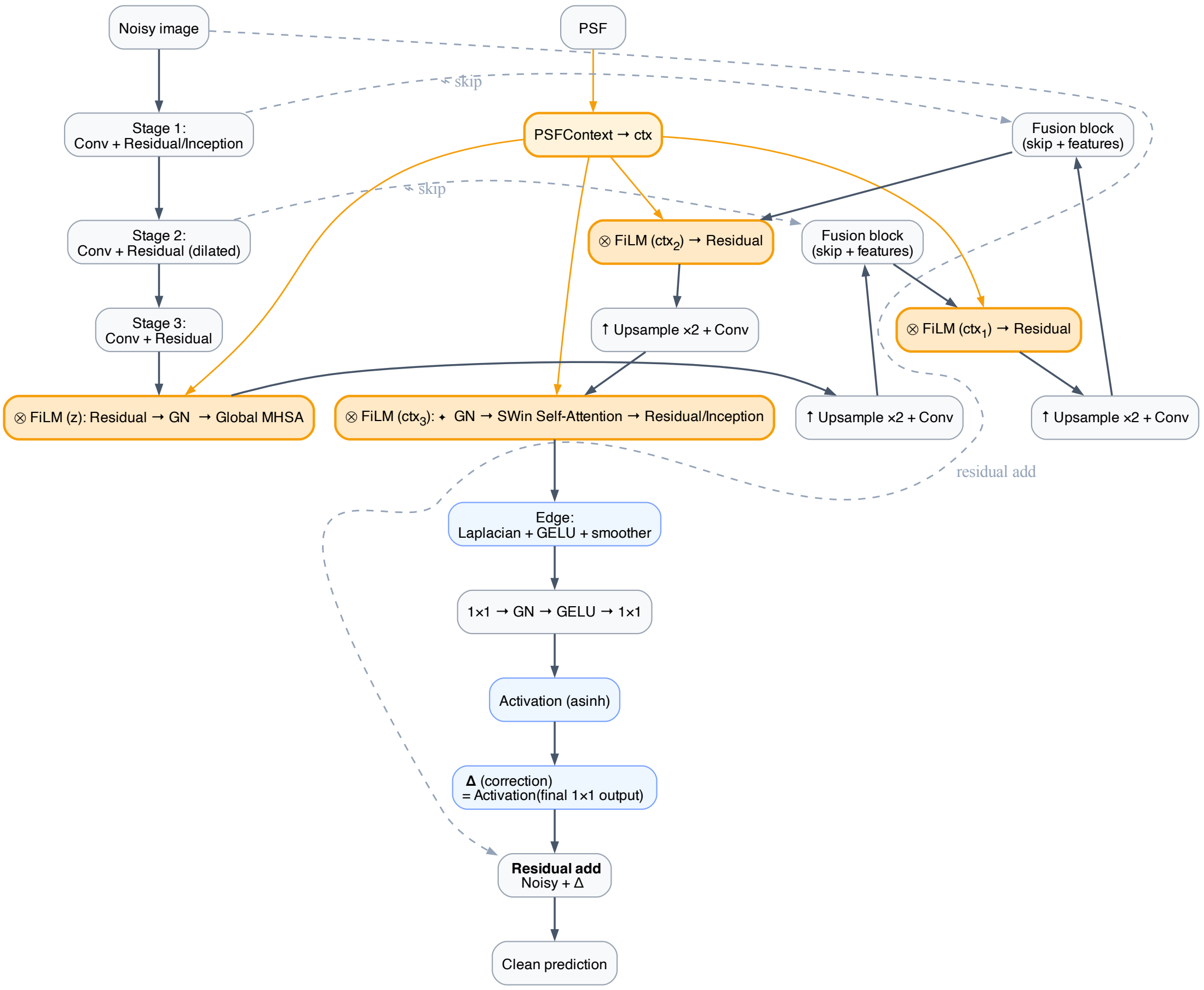

**Figure D1.** VIREO model architecture.

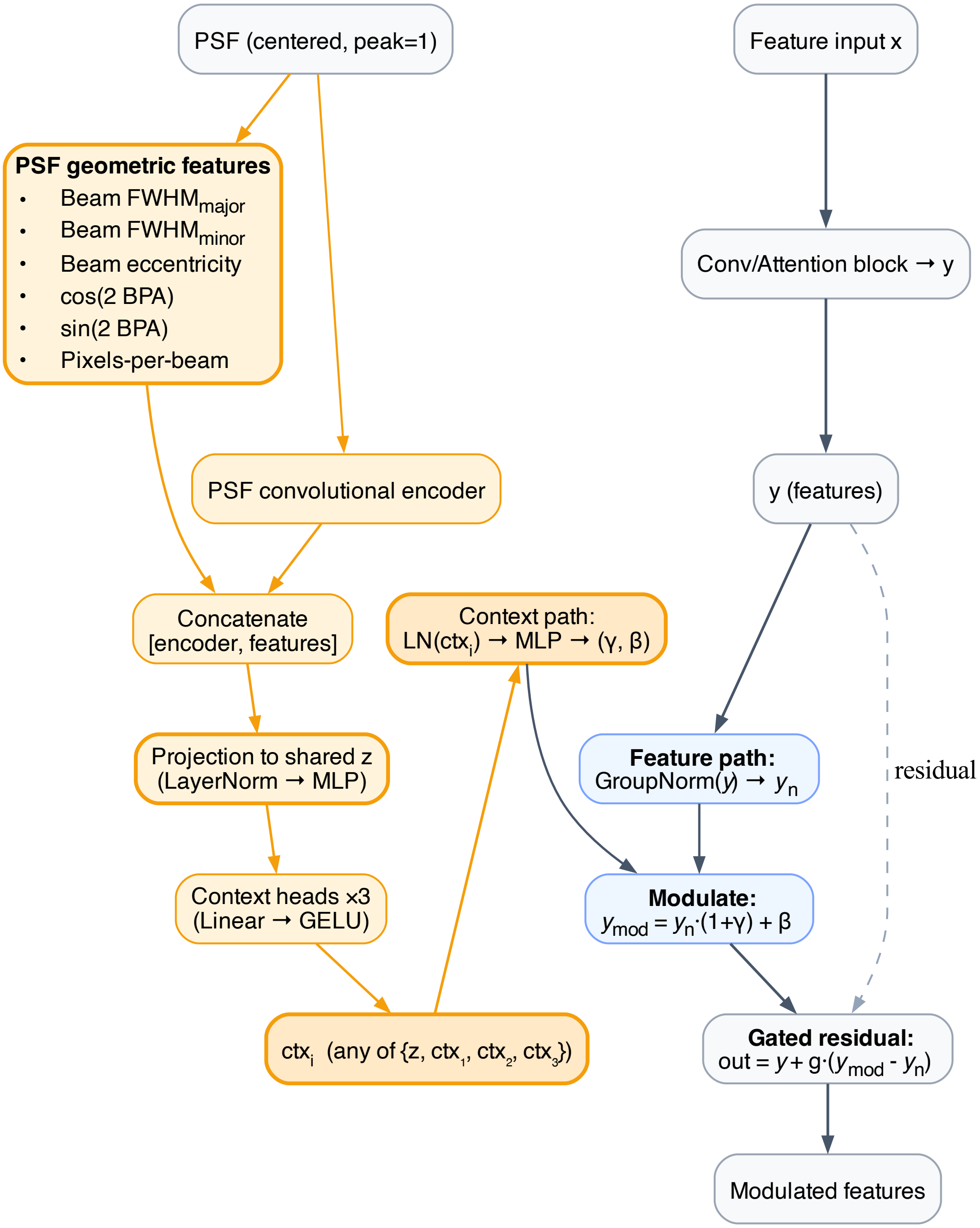

**Figure D2.** Architecture of FiLM layer.